\documentclass[journal]{IEEEtran}
\usepackage[normalem]{ulem}
\usepackage{dsfont}
\usepackage{amsmath,amssymb,amsfonts}
\usepackage{algorithmic}
\usepackage{graphicx}
\usepackage{textcomp}
\usepackage{caption}
\usepackage{subcaption}
\usepackage{latexsym}
\usepackage{amsmath}
\usepackage[mathscr]{euscript}
\usepackage[linesnumbered, algo2e, ruled,norelsize]{algorithm2e}
\SetArgSty{textnormal}
\usepackage{threeparttable}
\usepackage{threeparttablex}
\usepackage{commath}
\usepackage{mathtools}
\usepackage{subcaption}
\usepackage{slashbox}
\usepackage{pict2e}
\usepackage{epstopdf}
\usepackage{verbatim}
\usepackage{subfloat}
\usepackage{tabu}
\usepackage{booktabs}
\usepackage{float}
\usepackage{amssymb}
\usepackage{url}
\usepackage{multirow}
\usepackage{graphicx}
\usepackage{mathptmx}
\usepackage{cite}
\usepackage{footnote}
\usepackage{array}
\usepackage[table]{xcolor}
\usepackage{tabularx}
\usepackage{ulem}
\usepackage{algorithm}
\usepackage{amsthm} 
\def\BibTeX{{\rm B\kern-.05em{\sc i\kern-.025em b}\kern-.08em
    T\kern-.1667em\lower.7ex\hbox{E}\kern-.125emX}}

\DeclareMathOperator*{\argmaxA}{arg\,max} 

\begin{document}

\title{Multi-user Visible Light Communications with Probabilistic Constellation Shaping and Precoding}

\author{Thang~K.~Nguyen,~\IEEEmembership{Graduate Student Member,} Thanh~V.~Pham,~\IEEEmembership{Member,~IEEE,} Hoang~D.~Le, ~\IEEEmembership{Member,~IEEE,} Chuyen~T.~Nguyen 
        and Anh~T.~Pham,~\IEEEmembership{Senior Member,~IEEE}
\thanks{This work was supported by JSPS KAKENHI Grants Number 23K13333, 24K14918, and the Competitive Fund from the University of Aizu (P-6). Thang Nguyen is also supported by the Japanese government's Monbukagakusho (MEXT) Scholarship.

Thang~K.~Nguyen, Hoang~D.~Le, and Anh~T.~Pham are with the Department of Computer Science and Engineering, The University of Aizu, Fukushima, Japan (e-mail: m5272015@u-aizu.ac.jp, hoangle@u-aizu.ac.jp, pham@u-aizu.ac.jp). 

Thanh~V.~Pham is with the Department of Mathematical and Systems Engineering, Shizuoka University, Shizuoka, Japan (email: pham.van.thanh@shizuoka.ac.jp)

Chuyen~T.~Nguyen is with the School of Electrical and Electronic Engineering of Hanoi University of Science and Technology, Hanoi, Vietnam (email: chuyen.nguyenthanh@hust.edu.vn)

Part of this paper has been accepted for presentation at the 2024 IEEE Global Communications Conference (2024 IEEE GLOBECOM) \cite{Thang2024-Globecom}.
}} 
\maketitle
\begin{abstract}
This paper proposes a joint design of probabilistic constellation shaping (PCS) and precoding to enhance the sum-rate performance of multi-user visible light communications (VLC) broadcast channels subject to signal amplitude constraint. In the proposed design, the transmission probabilities of bipolar $M$-pulse amplitude modulation ($M$-PAM) symbols for each user and the transmit precoding matrix are jointly optimized to improve the sum-rate performance. The joint design problem is shown to be a complex multivariate non-convex problem due to the non-convexity of the objective function. To tackle the original non-convex optimization problem, the firefly algorithm (FA), a nature-inspired heuristic optimization approach, is employed to solve a local optima. The FA-based approach, however, suffers from high computational complexity. Thus, using zero-forcing (ZF) precoding, we propose a low-complexity design, which is solved using an alternating optimization approach. Additionally, considering the channel uncertainty, a robust design based on the concept of end-to-end learning with autoencoder (AE) is also presented. Simulation results reveal that the proposed joint design with PCS significantly improves the sum-rate performance compared to the conventional design with uniform signaling. For instance, the joint design achieves $\mathbf{17.5\%}$ and $\mathbf{19.2\%}$ higher sum-rate for 8-PAM and 16-PAM, respectively, at 60 dB peak amplitude-to-noise ratio. Some insights into the optimal symbol distributions of the two joint design approaches are also provided. Furthermore, our results show the advantage of the proposed robust design over the non-robust one under uncertain channel conditions.
\end{abstract}

\section{Introduction}
Over the past decade, the exponential growth of mobile devices and data-intensive multimedia applications has tremendously burdened current radio-frequency (RF) wireless systems. The immense demand for data traffic and high data-rate transmission has led to rapid progress in the research and development of new wireless technologies. Visible light communications (VLC), which leverages the visible light spectrum for data transmission, is emerging as a promising alternative or complement to existing RF technologies. With several unique advantages, such as providing high-capacity data transmission with a huge unlicensed spectrum and immunity to RF interference, VLC is expected to play a key role in the future ubiquitous networks\cite{VLC_survey_2019}. 

While input signals in RF communications can be complex and are often subject to average power constraints, input signals in VLC systems must be real, non-negative, and are constrained by a peak power (i.e., amplitude constraint) to ensure the limited linear range of the LEDs and/or to comply with the eye-safety regulations \cite{PLS_2014}. According to \cite{Capacity_peak_1971} and \cite{GM_2016}, the capacity-achieving input distribution for a scalar Gaussian channel under an amplitude constraint is discrete with a finite number of symmetric mass points. In practical systems with a particular modulation scheme, this implies that the position and transmission probability of the symbols should be jointly optimized to approach the channel capacity. In literature, the optimizations of symbols' position and transmission probability are known as geometric constellation shaping (GCS) and probabilistic constellation shaping (PCS), respectively. 
Using GCS, the positions of the constellation symbols are arranged to approximate the capacity-achieving input distribution \cite{GCS_2007, GCS_2013, GCS_2022}. However, GCS is typically impractical because finding the optimal locations of constellation points for arbitrary channel conditions is complex. Furthermore, the irregular arrangement of GCS constellation symbols increases transceiver complexity significantly and makes it difficult to maintain Gray mapping. Instead of arranging the constellation symbol's position, PCS shapes the probability of occurrence of constellation symbols, i.e., constellation symbols are chosen with a nonuniform probability distribution \cite{PCS_1993, PCS_2023, PCS_2024}. 
A probabilistically shaped constellation can be generated by a constant composition distribution matcher (CCDM) \cite{CCDM_2016}, which maps uniform information bits into symbols with the desired distribution. In contrast to GCS, constellation symbols in PCS are evenly spaced, enabling easy implementation of Gray mapping and not requiring upgrading or modifying the transceiver.  

Due to its low complexity and flexibility, PCS has been widely studied in optical fiber \cite{PCS_fiber_2019, rate_adaptation_PCS_2016, PCS_OFDM_fiber_2019} and optical wireless communications \cite{PCS_FSO_2020, PS_VLC_2022, OFDM_VLC_2020, PCS_secrecy_2024}. For optical fiber communication systems, the authors in \cite{PCS_fiber_2019} derived the optimal parameters of PCS and forward error correction (FEC) that maximize the information rate. In \cite{rate_adaptation_PCS_2016}, a rate adaptation system for single-carrier coherent optical transmission utilizing probabilistically shaped quadrature amplitude modulation (QAM) was proposed. And, in \cite{PCS_OFDM_fiber_2019}, a probabilistically shaped orthogonal frequency-division multiplexing (OFDM) modulation was proposed and experimentally demonstrated for optical access networks. For optical wireless communication, the authors in \cite{PCS_FSO_2020} proposed an adaptive modulation scheme based on PCS  to approach the capacity of the free-space optical (FSO) channels. With the same design objective, the authors in \cite{PS_VLC_2022} attempted to design a PCS-based spatial modulation for VLC channels. In \cite{OFDM_VLC_2020}, a VLC system using probabilistically shaped OFDM modulation was presented. In \cite{PCS_secrecy_2024}, the authors proposed a practical design of PCS for physical layer security (PLS) in VLC systems. However, it should be noted that the existing studies focused only on utilizing the PCS scheme in point-to-point (P2P) communications with one-user scenarios. 

Due to the broadcast nature of visible light signals, VLC systems can be categorized as broadcast networks. By exploiting the spatial degrees of freedom (DoF) of multiple light-emitting diode (LED) transmitters in the form of precoding, VLC systems have the capability to serve multiple users using the same time-frequency resource. However, the inherent multi-user interference (MUI) in MU broadcast systems can deteriorate the performance. Handling the MUI at the receivers is generally challenging without coordination among them. Therefore, it is more practical to mitigate the MUI at the transmitter side by proper precoding designs. The problem of precoding design has been investigated in several works considering different optimization objectives, for example, minimizing the sum mean square error (MSE) \cite{Ma2015} or the total transmit power \cite{Ma2019}, maximizing the sum-rate and fairness performance \cite{ZF_2017} with zero-forcing (ZF) precoding. 
It is important to note that, to facilitate the precoding design, previous works often considered a closed-form lower bound on the channel capacity derived by assuming the continuous uniform distribution of the input. As a consequence, the obtained results might not accurately reflect the actual system performance. Moreover, the shaping gain promised by PCS was also not taken into consideration. 

To address these two issues, it is essential to study precoding design considering the exact channel capacity of a particular modulation (i.e., modulation-constrained capacity) with PCS. 
In this regard, a joint optimization of precoding and PCS can simultaneously reduce the effect of MUI and approach channel capacity, hence improving the overall sum-rate performance. For the scenario of multiple LED transmitters serving a single user, the authors in \cite{joint_PCS_precoding_2022} proposed a joint PCS and precoding design to maximize the achievable rate under both peak and average amplitude constraints. The optimal design was handled by a two-step optimization procedure that sequentially solved the precoder and symbol distribution. For the case of a single user, these two optimization problems were shown to be convex and could be effectively solved. Unfortunately, in the multi-user broadcast system, the proposed approach can not be applied due to the presence of the MUI that destroys the convexity of the achievable rate formula (i.e., the objective function). Thus, solving a joint design of PCS and precoding for the multi-user VLC systems presents significant challenges as the optimal design problem is shown to be more complex and multivariate non-convex. To the best of our knowledge, no study has been done on this issue.

Against the above background, this paper introduces a joint design of precoding and PCS for multi-user VLC systems to maximize the sum-rate performance where the transmit constellation symbols are drawn from a probabilistically shaped $M$-ary pulse amplitude modulation ($M$-PAM). The design problem is shown to be multivariate non-convex, which renders solving the global optima challenging. Therefore, two sub-optimal design approaches are presented. The main contributions of this paper are speciﬁcally summarized as follows.
\begin{itemize}
    \item To address the non-convex problem, we present a firefly algorithm (FA) approach to solve a locally optimal solution to constellation probability distribution and precoding matrix simultaneously. Posing both exploitation and exploration abilities \cite{FA_2009, natural_meta_2014, Le2024}, FA is a promising candidate for solving multivariate non-convex design problems. It tends to be a global optimizer at the expense of computational complexity. 
    \item By adopting a particular precoding criterion, i.e., the zero-forcing (ZF) precoding, we propose a low-complexity joint design, which is solved using a combination of alternating optimization (AO) and successive convex approximation techniques (SCA).
    \item In practice, the assumption that the user's channel state information (CSI) is perfectly known at the transmitter is generally unrealistic, particularly when dealing with moving users. A robust design for maximizing the sum-rate given the channel uncertainty is therefore investigated. Due to the lack of information about actual channel vectors caused by imperfect CSI, the sum-rate maximization problem is intractable and difficult to solve explicitly using classical techniques such as convex optimization and meta-heuristic algorithms. We, thus, propose a robust design based on the end-to-end learning concept using autoencoder (AE). Through a robust training process, a robust design that achieves good performance over uncertain channel conditions can be obtained.   
    \item Comprehensive simulations are performed to evaluate the superiority of the proposed joint design over the precoding design with the conventional uniform distributed PAM. Some insights into the users' optimal symbol distributions under different values of peak amplitude-to-noise ratio are provided. In addition, our results indicate the advantage of the proposed robust design over the non-robust one when the channel uncertainty is taken into account. 
\end{itemize}

The balance of the paper is organized as follows. The system model is described in Section \ref{sec:model}. Section \ref{sec:problem} presents the joint design of PCS and precoding for maximizing the sum-rate. In Section \ref{sec:robust_AE}, taking into account the channel uncertainty, a robust design based on the end-to-end learning concept is proposed. Representative simulation results are given in Section \ref{sec:results}, and finally, Section \ref{sec:conclusion} concludes the paper.

\textit{Notation:} The following notations are used throughout the paper. $\mathbb{R}$ is the set of real-valued numbers. Bold upper case letters denote matrices, e.g., $\mathbf{A}$, whereas bold lower case letters indicate vectors, e.g., $\mathbf{a}$. The transpose of $\mathbf{A}$ is written as $\mathbf{A}^T$. The $i$-th row vector of matrix $\mathbf{A}$ is denoted as $\mathbf{[A]}_{i,:}$  and the $i$-th element of vector $\mathbf{a}$ are denoted as $\mathbf{[a]}_i$. $\mathbf{1}_{n}$ denote the all-ones vector of size $n$. Moreover, $\left\lVert.\right\rVert_1$, and $\left\lVert.\right\rVert_2$ respectively indicate the norm-1 and Euclidean norm.

\section{System Model}
\label{sec:model}
\begin{figure}[t]
    \centering
    \includegraphics[width = 8.8 cm]{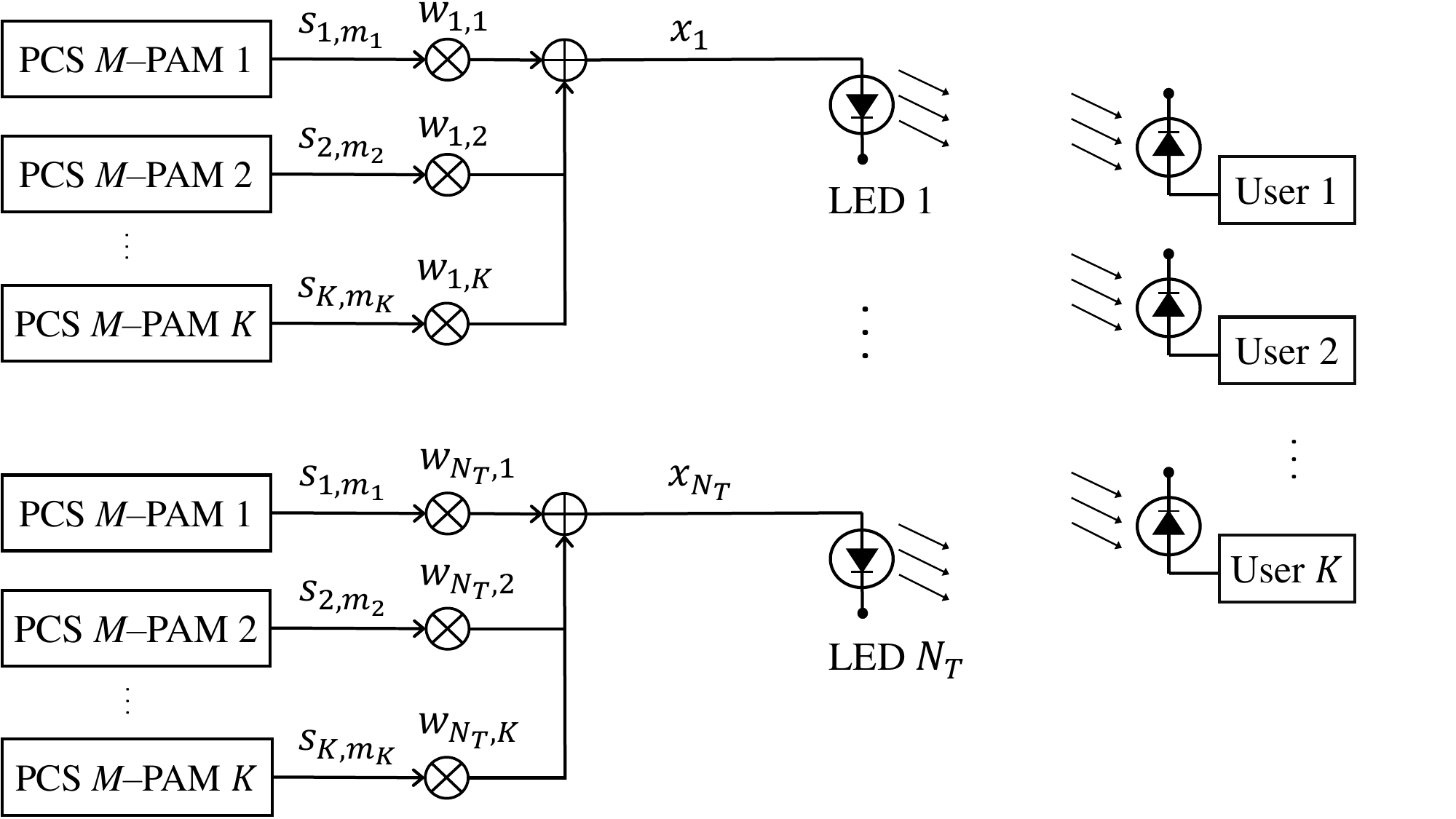}
    \caption{Multi-user VLC system with precoding and PCS.}
    \label{fig:system_model}
\end{figure}



Our considered multi-user VLC system, as illustrated via a simple example in Fig.~\ref{fig:system_model}, consists of $N_T$ LED luminaires, $K$ independent users where each user is equipped with a single photodiode (PD) receiver.
Let $\mathbf{s} = \begin{bmatrix}s_1 & s_2 & \cdots & s_K\end{bmatrix}^T \in \mathbb{R}^{K \times 1}$ be the vector of transmitted constellation symbols for $K$ users. The constellation symbol $s_k$ is generated from the $k$-th bipolar PCS $M$-PAM  modulator with $k = 1,\ 2, \ \cdots,\ K$.
For the $k$-th PCS $M$-PAM modulation, denote the set of $M$ equally spaced bipolar symbols as $\mathbb{S}_k = \{ s_{k,1}, \ s_{k,2}, \ \cdots, \ s_{k,M} \} = \{s_{k,m_k}\}_{\left(m_k = 1,\ 2,\ \cdots,\ M\right)}$ with the corresponding amplitude vector is  $\mathbf{a}_k = [ a_{k,1} \ a_{k,2} \ \cdots \ a_{k,M} ]= \begin{bmatrix}a_{k,m_k}\end{bmatrix}_{\left(m_k = 1,\ 2,\ \cdots,\ M\right)}$
and the corresponding probability mass function (PMF) vector is $\mathbf{p}_k = \begin{bmatrix} p_{k,1} \ p_{k,2} \ \cdots \ p_{k,M} \end{bmatrix}= \begin{bmatrix}p_{k,m_k}\end{bmatrix}_{\left(m_k = 1,\ 2,\ \cdots,\ M\right)}$. 
 
Denote $\mathbf{W} = \begin{bmatrix}\mathbf{w}_1 \ \mathbf{w}_2 \ \cdots \ \mathbf{w}_K \end{bmatrix}~ \in ~ \mathbb{R}^{N_T \times K}$ as the precoding matrix where $\mathbf{w}_k = \begin{bmatrix}w_{1,k} \ w_{2, k} \ \cdots \ w_{N_T, k}\end{bmatrix}^T$ is the $k$-th user's precoder vector with $w_{n,k}$ being the precoder of the transmitted signal from the $n$-th LED luminaire to $k$-th user for $n = 1,\ 2,\ \cdots, \ N_T$. Therefore, at the $n$-th LED transmitter, the broadcast signal $v_n$, which consists of constellation symbols of all users, can be given as
\begin{align}
\label{eqn: v_n}
    v_n = [\mathbf{W}]_{n,:} \times \mathbf{s},
\end{align}
where $[\mathbf{W}]_{n,:}$ is the $n$-th row of $\mathbf{W}$, which represents the precoder for $n$-th LED transmitter.
For illumination, a DC bias $I_{n}^{\text{DC}}$ is added to $x_n$ to generate a non-negative drive current $z_n$. In addition, to guarantee the operation of LEDs, i.e., to avoid the overheating problem and the potential light intensity reduction, $z_n$ must also be limited to a maximum threshold denoted by $I_{\text{max}}$. Therefore, we have 
\begin{align}
\label{eqn:peak_constraint1}
    0 \leq z_n=v_n+I_n^{\text{DC}} \leq I_{\text{max}},
\end{align}
From \eqref{eqn: v_n} and denote $A = \min\left\{I_n^{\text{DC}}, \ I_{\text{max}} - I_n^{\text{DC}}\right\}$, we have
\begin{align}
\label{eqn:peak_constraint2}
    -A \leq  [\mathbf{W}]_{n,:} \times \mathbf{s} \leq A.
\end{align}

Let us assume that for all PCS $M$-PAM modulators, the peak amplitudes of constellation symbols are the same and are denoted as $A$, i.e., $|a_{k,m_k}|~\leq~ A$. Due to the symbol symmetry around 0, for each PCS $M$-PAM modulation, the symbol amplitude levels can be given as $a_{k,m_k} = (2m_k-M-1)\frac{A}{M-1}$ for $m_k = 1,\ 2,\ \cdots, \ M$. 
Because the symbol amplitudes are in the range of $[-A,~A]$, to satisfy the peak amplitude constraint in \eqref{eqn:peak_constraint2}, the following constraint must be imposed
\begin{align}
    \left\lVert[\mathbf{W}]_{n,:}\right\rVert_1 \ \leq \ 1,\ \forall{n} \ \in \ \{1,\ 2,\ \cdots,\ N_T\}.
\end{align}

Let $\mathbf{H} = \begin{bmatrix} \mathbf{h}_1 & \mathbf{h}_2 & \cdots & \mathbf{h}_K\end{bmatrix}^{\text{T}}$  denote the channel matrix, where $\mathbf{h}_k = \begin{bmatrix} h_{1,k} & h_{2,k} & \cdots & h_{N_T,k}\end{bmatrix} ^\text{T} \in \mathbb{R}^{N_T \times 1}$ is the $k$-th user's channel vector with $h_{n,k}$ being the line-of-sight (LoS) channel coefficient between the $n$-th LED transmitter and the $k$-th user\footnote{Details on the VLC channel, which can be found extensively in the literature (for example, in \cite{Thang2024-Globecom}), are omitted here for brevity. }. At the $k$-th user, the received optical signals are captured by the PD and transformed into an electric signal as
\begin{align}
    \label{eqn:received_signal}
    y_k &= \mathbf{h}_k^T \mathbf{w}_k s_k + \underbrace{\mathbf{h}_k^T \sum_{i=1,\ i \neq k}^K \mathbf{w}_i s_i + n_k}_{\overline{y}_k},
\end{align}
where $n_k \ \sim \ \mathcal{N}(0, \ \sigma^2) $ is the additive white Gaussian noise (AWGN) with zero-mean and the power of $\sigma^2$. 


\section{Sum-rate maximization}
\label{sec:problem}
\subsection{Problem Formulation}
\begin{figure*}[h]
    \begin{align}
    \label{eqn:f_yk}
        f(y_k) = \sum_{p, \ (m_1, \ \cdots,\ m_K)_p \in \mathcal{A}} \left[\prod_{i = 1}^{K} \text{P}(s_i = s_{i,m_i}) \frac{1}{\sqrt{2\pi\sigma^2}} \exp{\left(-\frac{\left(y_k - \mathbf{h}^T_k \sum_{i=1}^K \mathbf{w}_i a_{i,m_i}\right)^2}{2\sigma^2}\right)}\right],
    \end{align}
    where $\mathcal{A} = \{m_1, \ \cdots,\ m_K\} \times\{1,\ \cdots,\ M\}$ is the Cartesian product of two sets and $(m_1, \ \cdots,\ m_K)_p$ is element $p$-th of set $\mathcal{A}$. 
    \begin{align}
    \label{eqn:f_yk'}
        f(\overline{y}_k) = \sum_{q, \ (m_1, \ \cdots,\ m_K)_q \in \mathcal{A}_k} \left[\prod_{j = 1, j \neq k}^{K} \text{P}(s_j = s_{j, m_{j}}) \times \frac{1}{\sqrt{2\pi\sigma^2}} \exp{\left(-\frac{\left(\overline{y}_k - \mathbf{h}^T_k \sum_{j=1,j\neq k}^K \mathbf{w}_j a_{j,m_j}\right)^2}{2\sigma^2}\right)}\right],
    \end{align}
    where $\mathcal{A}_k = \{\{m_1, \ \cdots,\ m_K\} \setminus{\{m_k\}\} \times\{1,\ \cdots,\ M\}}$ and $(m_1, \ \cdots,\ m_K)_q$ is element $q$-th of set $\mathcal{A}_k$. 
    \line(1,0){\linewidth}
\end{figure*}
The achievable rate of the $k$-th user can be given as
\begin{align}
    \label{eqn: R_k}
    R_{k} & = \mathbb{I}(s_k;y_k) = h(y_k) - h(y_k|s_k) = h(y_k)-h(\overline{y}_k), \\
          & = -\int_{-\infty}^{+\infty}f(y_k)\log_2{f(y_k)}\text{d}y_k + \int_{-\infty}^{+\infty}f(\overline{y}_k)\log_2{f(\overline{y}_k)}\text{d}\overline{y}_k, \nonumber
\end{align}
where $\mathbb{I}(\cdot; \cdot)$ is the mutual information, $h(\cdot)$ and $h(\cdot|\cdot)$ are the differential and conditional entropy, respectively.
$f(y_k)$ and $f(\overline{y}_k)$ are the probability density functions of $y_k$ and $\overline{y}_k$ and are respectively given by \eqref{eqn:f_yk} and \eqref{eqn:f_yk'}, which are on top of the next page.

Observe that when the constellation symbol amplitude levels in each PCS-PAM constellation i.e., $a_{k,m_k}$ are fixed, the achievable rate $R_k$ is a function of the precoding matrix $\mathbf{W}$ and symbol distribution vectors of $K$ PCS-PAM constellations, i.e., $\mathbf{p}_1,\ \mathbf{p}_2, \ \cdots, \ \mathbf{p}_K$. For the sake of mathematical analysis in the later parts of the paper, denote $\mathbf{P} = \begin{bmatrix}\mathbf{p}_1 & \mathbf{p}_2 & \cdots & \mathbf{p}_K\end{bmatrix}^T \in \mathbb{R}^{K \times M}$. 
Therefore, the sum-rate maximization problem can be formulated as
\begin{subequations}
    \label{original_problem}    
    \begin{alignat}{2}
        \mathbb{P} \mathbf{1}:~
        &\underset{\mathbf{P}, \mathbf{W}}{\text{maximize}} \ \ \sum_{k=1}^K R_k (\mathbf{P}, \mathbf{W})
        \label{eqn: objectiveP1}\\ 
        &\text{subject to } & & \nonumber \\       
        & \hspace{1 cm} \left\lVert[\mathbf{W}]_{n,:}\right\rVert_1 \ \leq \ 1,\label{eqn:constraint_peak}\\
        & \hspace{1 cm} 0 \ \leq \ \mathbf{P}  \leq \ 1 , \label{eqn:constraint_P1}\\
        & \hspace{1 cm}  \mathbf{P} \times \mathbf{1}_{M} = \mathbf{1}_{K},\label{eqn:constraint_P2}
    \end{alignat}    
\end{subequations}
where \eqref{eqn: objectiveP1} is the sum rate formula and \eqref{eqn:constraint_peak} is the peak power constraint for LEDs. It can be observed that problem $\mathbb{P} \mathbf{1}$ is non-convex due to the non-convexity of the objective function. Thus, it is generally difficult to optimally solve it. 
Motivated by the above observations, we introduce a novel Firefly algorithm (FA) approach to simultaneously solve $\mathbf{P}$ and $\mathbf{W}$ for the original $\mathbb{P} \mathbf{1}$ in the following sections.


\subsection{Proposed Firefly Algorithm}
\subsubsection{Firefly Algorithm Implementation}
Adopting the penalty method as \cite{natural_meta_2014}, problem $\mathbb{P} \mathbf{1}$ can be equivalently reformulated as
\begin{align}
    \mathbb{P} \mathbf{2}:~
    &\underset{\mathbf{P}, \mathbf{W}}{\text{maximize}} \ \ \sum_{k=1}^K R_k(\mathbf{P}, \mathbf{W})  - P(\mathbf{P}, \mathbf{W}),
    \label{objectiveP2}
\end{align}
where $P(\mathbf{P}, \mathbf{W})$ is the penalty term, which is given as 
\begin{align}
    \label{eqn:penalty_term}
    &P(\mathbf{P}, \mathbf{W}) = 
    \lambda_1 \sum_{n=1}^{N_t}\max{\left(0, ~\left\lVert[\mathbf{W}]_{n,:}\right\rVert_1 - 1\right)}^2\\\nonumber
    & + \lambda_2 \sum_{k=1}^{K}\sum_{m=1}^M \min{\left(0,~ p_{k,m}\right)}^2 + \lambda_3 \sum_{k=1}^{K}\sum_{m=1}^M \max{\left(0,~ p_{k,m} - 1\right)}^2 \\\nonumber
    & + \lambda_4 \sum_{k=1}^K \max{\left(0, ~\left\lVert[\mathbf{P}]_{k,:}\right\rVert_1 - 1\right)}^2,
\end{align}
where $\lambda_j$ are penalty constants. The FA was proposed based on the firefly behaviors with three idealized rules \cite{FA_2009}, \cite{natural_meta_2014}. First, all fireflies are unisex, so one firefly will be attracted to other fireflies regardless of sex. Second, the attractiveness of any firefly to the other one is proportional to its brightness, and both decrease as their distance increases. For any two flashing fireflies, the less bright one will move towards the brighter one. If there is no brighter one than a particular firefly, it will move randomly. Third, the brightness of a firefly is determined by the landscape of the objective function. 

Let $(\mathbf{W}_n, \mathbf{P}_n)$ be the particular location of $n$-th firefly  amongst the population of $N$ fireflies, i.e., $n \in \{1,2, \cdots, N\}$. Since the proposed optimization problem is a maximization, the brightness of the $n$-th firefly is determined as the value of the objective function at $(\mathbf{W}_n, \mathbf{P}_n)$ as
\begin{align}
    \label{eqn:light_intensity}
    I(\mathbf{W}_n, \mathbf{P}_n) = \sum_{k=1}^K R_k(\mathbf{W}_n, \mathbf{P}_n)  - P(\mathbf{W}_n, \mathbf{P}_n).
\end{align}
For any $m$-th and $n$-th fireflies among the population in the generation $t$, if $I\left(\mathbf{W}^{(t)}_n, \mathbf{P}^{(t)}_n\right) > I\left(\mathbf{W}^{(t)}_m, \mathbf{P}^{(t)}_m\right)$, the $m$-th firefly  will move toward the $n$-th firefly as
\begin{align}
    \label{eqn:update_W}
    &\mathbf{W}_m^{(t+1)} \!=\! \mathbf{W}_m^{(t)} \!+\! \beta_0 \exp{\left(\!-\!\gamma\left(r^{(t)}_{\mathbf{W},mn}\right)^2\right)} \left(\mathbf{W}^{(t)}_n \!-\! \mathbf{W}^{(t)}_m\right) \!+\! \zeta^{(t)} \mathbf{V_W},\\
    &\mathbf{P}_m^{(t+1)} \!=\! \mathbf{P}_m^{(t)} \!+\! \beta_0 \exp{\left(\!-\!\gamma\left(r^{(t)}_{\mathbf{P},mn}\right)^2\right)} \left(\!\mathbf{P}^{(t)}_n \!-\! \mathbf{P}^{(t)}_m\!\right) \!+\! \zeta^{(t)} \mathbf{V_P}, \label{eqn:update_P}
\end{align}
where $r^{(t)}_{\mathbf{W},mn} = \left\lVert \mathbf{W}^{(t)}_n - \mathbf{W}^{(t)}_m\right\rVert_2$ and $r^{(t)}_{\mathbf{P},mn} = \left\lVert \mathbf{P}^{(t)}_n - \mathbf{P}^{(t)}_m\right\rVert_2$, $\beta_0$ is the attractiveness at $r^{(t)}_{\mathbf{W},mn} = 0$, $r^{(t)}_{\mathbf{P},mn} = 0$ and $\gamma$ is the variation of attractiveness. The second terms in the right-hand side of  
\eqref{eqn:update_W} and \eqref{eqn:update_P} capture the attractions with $\beta = \beta_0 \exp{\left(-\gamma r^2\right)}$ is the attractiveness. The third terms in \eqref{eqn:update_W} and \eqref{eqn:update_P} are randomization with 
$\zeta^{(t)} = \zeta_0^t$ is the random factor at generation $t$, $\zeta_0$ is the initial random factor, and $\mathbf{V_W} ~ \in ~\mathbb{R}^{N_T \times K}$, $\mathbf{V_P} ~ \in ~\mathbb{R}^{K \times M}$ are random matrices whose elements are drawn from a normal distribution. The proposed FA is summarized in \textbf{Algorithm \ref{alg1}}. 
\begin{algorithm}[ht]
    \caption{ FA for solving $\mathbb{P} \mathbf{2}$}\label{alg1}
    \begin{algorithmic}[1]
    \STATE \textbf{Input}: Population size $N$, maximum generation $T$.
    \STATE Generate $N$ populations $\{\left(\mathbf{W}_1, \mathbf{P}_1\right),  \cdots, \left(\mathbf{W}_N, \textbf{P}_N\right)\}$ randomly.
    \STATE Evaluate the light intensities of $N$ population $I(\mathbf{W}_i, \mathbf{P}_i) ~ \forall{i ~ \in ~ [1, \ N]}$ as \eqref{eqn:light_intensity}.
    \STATE Rank the fireflies in descending order of light intensities $I(\mathbf{W}_i, \mathbf{P}_i)$.
    \STATE Define the current best solution: \\
    $(\mathbf{W}^*,\mathbf{P}^*) \leftarrow (\mathbf{W}_1, \mathbf{P}_1), \ I^* \leftarrow I(\mathbf{W}^*, \mathbf{P}^*)$.
            \FOR{$t = 1~:~T$}
                \FOR{$m = 1~:~N$} 
                    \FOR{$n = 1~:~N$} 
                        \IF{$I(\mathbf{W}_n, \mathbf{P}_n) > I(\mathbf{W}_m, \mathbf{P}_m)$} 
                             \STATE 1. Move the $m$-th firefly toward the $n$-th firefly described in \eqref{eqn:update_W}, \eqref{eqn:update_P}.
                             \STATE 2. Update the light intensity of the $m$-th firefly with new $\left(\mathbf{W}_m, \mathbf{P}_m\right)$ as \eqref{eqn:light_intensity}.
                        \ENDIF
                    \ENDFOR
                \ENDFOR 
                \STATE Rank the fireflies in descending order of $I(\mathbf{W}_i, \mathbf{P}_i)$.
                \STATE Update the current best solution $(\mathbf{W}^*, \mathbf{P}^*) \leftarrow (\mathbf{W}_1, \mathbf{P}_1), \ I^* \leftarrow I(\mathbf{W}^*, \mathbf{P}^*)$.
            \ENDFOR
    \STATE Return the solution $(\mathbf{W}^*, \mathbf{P}^*)$.
    \label{alg.1}
    \end{algorithmic}
\end{algorithm}

\subsubsection{Asymptotic Optimality and Convergence}
As a metaheuristic algorithm, the convergence of the FA, like many other nature-inspired algorithms, has yet to be rigorously proven despite many applications.
In this section, following the same argument in \cite{FA_2009} and \cite{Le2024}, an analysis for the optimality and convergence of the FA is provided specifically for the considered problem in our paper.

Considering two limiting cases when the variation of attractiveness $\gamma \rightarrow 0$ and $\gamma \rightarrow \infty$. When $\gamma \rightarrow 0$, $\exp{\left(-\gamma r^2_{\mathbf{W}, mn}\right)} \rightarrow 1$ and $\exp{\left(-\gamma r^2_{\mathbf{P}, mn}\right)} \rightarrow 1$, the attractiveness in \eqref{eqn:update_W} and \eqref{eqn:update_P} are constant and equal to $\beta_0$. It is equivalent to an idealized sky scenario where every firefly's light intensity does not decrease over distance, and each flashing firefly can be seen everywhere. Consequently, a global optimum can easily be achieved. The convergence of FA in this case is similar to that of Particle Swarm Optimization (PSO), which was analyzed in \cite{PSO_convergence_2002}. On the other hand, when $\gamma \rightarrow \infty$, $\exp{\left(-\gamma r^2_{\mathbf{W}, mn}\right)} \rightarrow 0$ and $\exp{\left(-\gamma r^2_{\mathbf{P}, mn}\right)} \rightarrow 0$, which indicates that the attractiveness of each firefly is almost zero in the sight of other fireflies. It is equivalent to a heavily foggy region where each firefly roams randomly and can not be seen by other fireflies. This corresponds to the completely random search approach, and the optimality is not always guaranteed. Although FA behaves like a random search in this limiting case, its solution perturbation or modification is similar to that of Simulated Annealing (SA). In \cite{SA_1993}, the SA has been demonstrated to be convergent under appropriate cooling conditions. For FA, the reduction of the roaming randomness $\zeta$ can be seen as a type of cooling schedule. Therefore, it can be anticipated that FA will converge in this case.

In fact, for FA, the attractiveness is usually between these two extremes, i.e., $0 < \gamma < +\infty$. By adjusting the attractiveness variation $\gamma$ and roaming randomness $\alpha_0$, FA can effectively find the global optima and all the local optima simultaneously and outperforms both the random search and particle swarm optimization (PSO) \cite{FA_2009}, \cite{natural_meta_2014}. Given a very large firefly population $N$, and assuming that $N$ significantly exceeds the number of local optima, the initial positions of these fireflies should be uniformly spread across the entire search space. As the iterations of \textbf{Algorithm 1} progress, i.e., $t$ increases, these fireflies will gradually converge to all locally brighter spots, including both local and global optima, in a stochastic manner. The global optimum can be determined by evaluating and comparing the brightest fireflies among these locally bright areas (i.e., the best solutions among the local optima). Theoretically, FA can approach global optima when $n \rightarrow \infty$ and $t \gg 1$. However, in \cite{FA_2009, natural_meta_2014, Le2024}, FA is reportedly converged typically with less than 50 to 100 generations. It tends to be a global optimizer but at the potential expense of large computational complexity. 


\subsection{Low-complexity Design with Zero-forcing Precoding}
To avoid the computational complexity of the joint design probabilistic shaping and precoding based on the FA approach, we employ the suboptimal zero-forcing (ZF) precoding strategy. 
Although suboptimal, ZF precoding performs well at the high signal-to-noise ratio (SNR) regime, which is typically realizable in VLC systems.  

With ZF precoding, the MUI at each user's received signal $y_k$ is completely removed via the construction of precoder $\mathbf{w}_k$ in such a way that is orthogonal to the channel vectors of other users, i.e., $\mathbf{h}^T_k \mathbf{w}_i = 0, \ \forall{i \neq k}$ \cite{ZF_2017}. This orthogonality results in lower computational complexity at the expense of decreased performance since the degrees of freedom in designing $\mathbf{W}$ is reduced compared to the general case.  
By removing the MUI via ZF precoding, the received electrical signal at user $k$-th is simplified to
\begin{align}
\label{eqn: received_signal_ZF}
    y_k =  \mathbf{h}_k^T \mathbf{w}_k s_k + n_k.
\end{align}
As a result, the achievable rate of the $k$-th user is given by
\begin{align}
    R_{k}(\mathbf{P}, \mathbf{W}) & = \mathbb{I}(s_k;y_k)  
            = h(y_k) - h(y_k|s_k) = h(y_k) - h(n_k) \\\nonumber
          & = -\int_{-\infty}^{+\infty}f(y_k)\log_2{f(y_k)}\text{d}y_k -\frac{1}{2}\log_2(2\pi e \sigma^2),
\end{align}
where $f(y_k)$ is the probability density function of $y_k$
\begin{align}
    f(y_k) =\sum_{m_k = 1}^{M} p_{k,m_k} \frac{1}{\sqrt{2\pi\sigma^2}} \exp{\left(\!\!\!-\frac{\left(y_k - \mathbf{h}^T_k \mathbf{w}_k a_{k,m_k}\right)^2}{2\sigma^2}\right)},  
\end{align}
with $m_k = 1,\ 2,\ \cdots,\ M$ and $p_{k,m} = \text{Pr}(s_k = s_{k,m_k})$.
By quantizing the continuous source $y_k$ into $N$ discrete values with a sufficiently small step size $\Delta$, the achievable rate $R_k(\mathbf{P}, \mathbf{W})$ can be alternatively derived by a Riemann sum with $N$ rectangular partitions whose width is $\Delta$ and height is $f(y_k^n)\log_2{f(y_k^n)}$
\begin{align}
    R_{k}&(\mathbf{P}, \mathbf{W}) = -\sum_{n=1}^N f(y_k^n)\log_2{f(y_k^n)}\Delta -\frac{1}{2}\log_2(2\pi e \sigma^2)\nonumber \\
    &= -\sum_{n=1}^N \left(\frac{1}{\sqrt{2\pi \sigma^2}} \sum_{m_k}^M p_{k,m_k} \exp{\left(-\frac{\left(y_k^n - \mathbf{h}_k^T \mathbf{w}_k a_{k,m_k}\right)^2}{2\sigma^2}\right)}\right)\nonumber \\
    &\times \log_2{\left(\frac{1}{\sqrt{2\pi \sigma^2}} \sum_{m_k}^M p_{k,m_k} \exp{\left(-\frac{\left(y_k^n - \mathbf{h}_k^T \mathbf{w}_k a_{k,m_k}\right)^2}{2\sigma^2}\right)}\right)}\Delta \nonumber \\
    & - \frac{1}{2}\log_2(2\pi e \sigma^2).
\end{align}
Then, the sum-rate maximization problem in the case of using ZF precoding can be formulated as
\begin{subequations}
    \label{ZF_problem}    
    \begin{alignat}{2}
        \mathbb{P} \mathbf{3}:~
        &\underset{\mathbf{P}, \mathbf{W}}{\text{maximize}} \ \ \sum_{k=1}^K R_k (\mathbf{P}, \mathbf{W})
        \label{eqn: objectiveP3}\\ 
        &\text{subject to } & & \nonumber \\       
        & \hspace{1 cm} \mathbf{h}^T_k \mathbf{w}_i = 0, \ \ \forall{i \neq k},\label{eqn:constraint_ZF}\\
        & \hspace{1 cm} \eqref{eqn:constraint_peak},~\eqref{eqn:constraint_P1},~\eqref{eqn:constraint_P2}. \nonumber
    \end{alignat}    
\end{subequations}

Due to the non-convexity of the objective function, $\mathbb{P} \mathbf{3}$ is a non-convex 
problem with two optimization variables $\mathbf{P}$ and $\mathbf{W}$. To tackle the problem, as $\mathbf{P}$ and $\mathbf{W}$ are two independent variables, they can be alternatively solved with an alternating optimization (AO) approach \cite{AO_2008}, which involves an iterative procedure where in each iteration one variable is optimized given the other is fixed. This process is repeated until a convergence criterion is met. 
Specifically, the procedure of the AO approach for solving \eqref{ZF_problem} is presented as follows.

1) Firstly, starting with an initial value of precoding matrix $\mathbf{W}^{(0)}$, at the $r$-th iteration of the procedure, the following sub-problem with respect to the PMF matrix  $\mathbf{P}$ is solved
    \begin{subequations}
    \label{ZF_problem_P}    
        \begin{alignat}{2}
            \mathbb{P} \mathbf{3(a)}:~
            &\underset{\mathbf{P}}{\text{maximize}} \ \ \sum_{k=1}^K R_k (\mathbf{P}, \mathbf{W}^{(r-1)})
            \label{eqn: objectiveP3a}\\ 
            &\text{subject to } & & \nonumber \\  
            & \hspace{1 cm} \eqref{eqn:constraint_P1},~\eqref{eqn:constraint_P2}. \nonumber
        \end{alignat}    
    \end{subequations}
    Here, $\mathbf{W}^{(r-1)}$ is the solution of $\mathbf{W}$ at the previous iteration. It is seen that $\mathbb{P} \mathbf{3(a)}$ is convex since the objective function is a concave function of $\mathbf{P}$ according to \cite[Theorem 2.7.4]{Elements_IT_2012} and the two constraints are linear. Therefore, it can be solved efficiently using standard optimization packages, such as CVX \cite{cvx}. 
    
2) Given the solution to $\mathbf{P}$ at the $r$-th iteration, i.e., $\mathbf{P}^{(r)}$, the precoding matrix $\mathbf{W}^{(r)}$ is then updated by solving the following sub-problem
    \begin{subequations}
    \label{ZF_problem_W}    
        \begin{alignat}{2}
            \mathbb{P} \mathbf{3(b)}:~
            &\underset{\mathbf{W}}{\text{maximize}} \ \ \sum_{k=1}^K R_k (\mathbf{P}^{(r)}, \mathbf{W})
            \label{eqn: objectiveP3b}\\ 
            &\text{subject to } & & \nonumber \\   
            & \hspace{1 cm} \eqref{eqn:constraint_ZF},~ \eqref{eqn:constraint_peak}. \nonumber
        \end{alignat}    
    \end{subequations}
    With respect to $\mathbf{W}$, it is observed that $\mathbb{P} \mathbf{3(b)}$ is not convex due to the non-convexity of the objective function.
    To tackle it, the successive convex approximation (SCA) is employed to solve local optima. To this end, by introducing the  slack variables $x_k^n$, $\forall{k = 1,\ \cdots, \ K}$ and $\forall{n = 1, \ \cdots, \ N}$, $\mathbb{P} \mathbf{3(b)}$ can be reformulated as
    \begin{subequations}
    \label{ZF_problem_W2}    
        \begin{alignat}{2}
            \label{eqn: objectiveP3b2}
            &\overline{\mathbb{P} \mathbf{3(b)}}:~ \nonumber
            \underset{\mathbf{W}, \ x_k^n}{\text{maximize}} \\
            & \sum_{k=1}^K \sum_{n=1}^N \left(\frac{-1}{\sqrt{2\pi \sigma^2}} \sum_{m=1}^M p_{k,m}^{(r)} x_k^n\right) \log_2{\left(\frac{1}{\sqrt{2\pi \sigma^2}} \sum_{m=1}^M p_{k,m}^{(r)} x_k^n\right)}\Delta  \\\nonumber
            & - \frac{1}{2}\log_2(2\pi e \sigma^2) \\\nonumber
            &\text{subject to } & & \nonumber \\       
            & \label{eqn:constraint_slack_var2} \hspace{1 cm}x_k^n = \exp{\left(-\frac{\left(y_k^n - \mathbf{h}_k^T \mathbf{w}_k a_{k,m_k}\right)^2}{2\sigma^2}\right)}, \\\nonumber  
            & \hspace{1 cm} \eqref{eqn:constraint_ZF},~ \eqref{eqn:constraint_peak}.
        \end{alignat}    
    \end{subequations}
    It is seen that the objective function is concave and constraints \eqref{eqn:constraint_ZF} and \eqref{eqn:constraint_peak} are convex, but \eqref{eqn:constraint_slack_var2} is not. To cope with this issue, the first-order Taylor approximation is employed to obtain a linear approximation of the non-convex constraint. Accordingly, at the $j$-th iteration, the SCA involves solving the following problem
    \begin{subequations}
    \label{ZF_problem_W3}    
        \begin{alignat}{2}
            \label{eqn: objectiveP3b3}
            &\widehat{\mathbb{P} \mathbf{3(b)}}:~ \nonumber
            \underset{\mathbf{W}, \ x_k^n}{\text{maximize}} \hspace{0.5 cm}\eqref{eqn: objectiveP3b2} \\\nonumber
            & \text{subject to } & &  \\    
            & \hspace{1 cm} x_k^n = 
            \exp{\left(-\frac{z_{\{n, j-1\}}^2}{2\sigma^2}\right)} \nonumber \\
            & \hspace{1 cm} + \exp{\left(-\frac{z_{\{n, j-1\}}^2}{2\sigma^2}\right)} \frac{z_{\{n, j-1\}}}{\sigma^2} \mathbf{h}_k^T \left(\mathbf{w}_k - \mathbf{w}_k^{(j-1)}\right)a_{k,m_k},\\\label{eqn:constraint_slack_var3}  
            & \hspace{1 cm} \eqref{eqn:constraint_ZF},~ \eqref{eqn:constraint_peak}, \nonumber 
        \end{alignat}    
    \end{subequations}
    where $z_{\{n, j-1\}} = y_k^n - \mathbf{h}_k^T \mathbf{w}_k^{(j-1)} a_{k,m_k}$, and $\mathbf{w}_k^{(j-1)}$ is the solution of $\mathbf{w}_k$ at the $(j-1)$-th iteration of the algorithm. Problem $\widehat{\mathbb{P} \mathbf{3(b)}}$ is now convex and can be solved using CVX. The SCA algorithm for solving $\overline{\mathbb{P} \mathbf{3(b)}}$ is described in \textbf{Algorithm 2}. 
\begin{algorithm}[ht]
\label{alg2}
    \caption{SCA algorithm for solving $\overline{\mathbb{P} \mathbf{3(b)}}$}
    \begin{algorithmic}[1]
    \STATE \textbf{Input}: Maximum number of iterations ${N}_{\text{max}}$, the error tolerance $\epsilon$. 
    \STATE Generate a feasible starting point $\mathbf{W}^{(0)}$.
    \WHILE{convergence = \textbf{False} and $j \leq {N}_{\text{max}}$} 
    \STATE Solve $\widehat{\mathbb{P} \mathbf{3(b)}}$ using $\mathbf{W}^{(j-1)}$ from the $(j-1)$-th iteration. 
    \STATE \IF{$\frac{\norm{\mathbf{W}^{(j)} - \mathbf{W}^{(j-1)}}}{\norm{\mathbf{W}^{(j)}}}\leq \epsilon$} 
    \STATE convergence = \textbf{True}
    \STATE $\mathbf{W}^* \leftarrow \mathbf{W}^{(j)}$ \ELSE
    \STATE convergence = \textbf{False} \ENDIF
    \ENDWHILE
    \STATE $j \leftarrow j+1 $
    \STATE Return the solution $\mathbf{W}^*$.
    \end{algorithmic}
\end{algorithm}

The AO approach repeatedly solves $\mathbb{P} \mathbf{3(a)}$ and $\mathbb{P} \mathbf{3(b)}$ in $N_0$ iterations to obtain a solution to $\mathbb{P} \mathbf{3}$.

\section{Robust autoencoder design with channel uncertainty}
\label{sec:robust_AE}
The assumption that the users' CSI are perfectly known at the transmitter is not always unrealistic, particularly when dealing with moving users. In RF communications, the transmitter can estimate the CSI using uplink-downlink reciprocity. In contrast, for VLC, the CSI must be estimated at the receiver and is fed back to the transmitter via an RF or infrared uplink. In the case of quick user movement, this process may result in outdated CSI estimations. Consequently, we propose in this section a robust joint design of PCS and precoding, considering the channel uncertainties. 

\subsection{Problem Formulation}
\label{sec:4A}
Suppose the CSI acquisition at the transmitter is imperfect, we consider an additive uncertainty in the channel vector
as 
\begin{align}
    \mathbf{h}_k = \widehat{\mathbf{h}}_k + \mathbf{u}_k,~~ \forall k = 1, \ 2, \cdots,\ K, 
\end{align}
where $\widehat{\mathbf{h}}_k$ is an estimation of the actual channel vector $\mathbf{h}_k$ and $\mathbf{u}_k$ is the CSI error. Note that $\widehat{\mathbf{h}}_k$ is known to the transmitter via outdated feedback from the $k$-th user. In the case of outdated CSI caused by user movement, the degree of CSI error can be characterized by a norm-bounded model as \cite{robust_BF_2015}, \cite{robust_EE_2021}
\begin{align}
    \left\lVert \mathbf{u}_k\right\rVert_2 \ \leq \ \delta_k, 
    \label{norm-bounded-model}
\end{align}
where $\delta_k$ represents the maximal changing level between the estimated and actual channels. While providing an explicit characterization of $\delta_k$ is challenging, considering that the bounded CSI error could be closely related to the instantaneous channel vector as studied in \cite{robust_BF_2015}, a simplified model for the CSI error level $\delta_k$ can be given by
\begin{align}
    \delta_k = \alpha \left\lVert \widehat{\mathbf{h}}_k \right\rVert_2,
\end{align}
where $\alpha \ \in [0,1)$ is a constant for evaluating the magnitude of CSI error. With the above-defined channel uncertainty model, the received signal $y_k$ at the $k$-th user is given as
\begin{align}
    \label{eqn:received_signal_uncertainty}
   \Tilde{y}_k &= \left(\widehat{\mathbf{h}}_k^T + \mathbf{u}_k^T\right) \mathbf{w}_k s_k + \underbrace{\left(\widehat{\mathbf{h}}_k^T + \mathbf{u}_k^T\right) \sum_{i=1,\ i \neq k}^K \mathbf{w}_i s_i + n_k}_{\hat{y}_k}.
\end{align}

Similar to the derivation in \eqref{eqn: R_k}, the achievable rate of the $k$-th user considering the channel uncertainty is expressed by
\begin{align}
\label{eqn:R_k_uncertainty}
    &\Tilde{R}_{k}(\mathbf{P}, \mathbf{W})  = \mathbb{I}(s_k;\Tilde{y}_k) \\\nonumber
          & = -\int_{-\infty}^{+\infty}f(\Tilde{y}_k)\log_2{f(\Tilde{y}_k)}\text{d}\Tilde{y}_k + \int_{-\infty}^{+\infty}f(\hat{y}_k)\log_2{f(\hat{y}_k)}\text{d}\hat{y}_k.
\end{align}
A joint design of PCS and precoding that is robust to channel uncertainty is thus formulated as  
\begin{subequations}
    \label{robust_problem}    
    \begin{alignat}{2}
        \mathbb{P} \mathbf{4}:~
        &\underset{\mathbf{P}, \mathbf{W}}{\text{maximize}} \ \ \sum_{k=1}^K \Tilde{R}_k (\mathbf{P}, \mathbf{W})
        \label{eqn: objectiveP_robust}\\ 
        &\text{subject to } & & \nonumber \\ 
        & \hspace{1 cm} \eqref{eqn:constraint_peak},~\eqref{eqn:constraint_P1},~\eqref{eqn:constraint_P2}. \nonumber
    \end{alignat}    
\end{subequations}
Note that in the case of the imperfect CSI, the transmitter only knows the estimated channel $\widehat{\mathbf{h}}_k$ and the norm bound of CSI error $\delta_k$. Moreover, due to the norm-bounded model in \eqref{norm-bounded-model}, there are infinite realizations of the CSI error $\mathbf{u}_k$ that renders solving $\mathbb{P} \mathbf{4}$ challenging. 
In the literature, one common approach to deal with channel uncertainty is to derive a lower bound on the achievable rate, which is easier to handle \cite{Hsiao2019,Liu2020}. Then, a worst-case design based on the derived bound is considered. It should be noted, however, that while derivation of such a lower bound is possible given simple closed-form achievable rate expressions in these previous studies, it seems not the case for the integral formula in \eqref{eqn:R_k_uncertainty}.      
The difficulty of solving $\mathbb{P} \mathbf{4}$ by traditional techniques such as meta-heuristic algorithms or convex optimization motivates us to examine the use of deep learning (DL) techniques. Specifically, based on the concept of end-to-end learning using AE, we propose a robust joint design 
to combat the uncertain channel conditions through a robust training process. 

\begin{figure*}[ht]
    \centering
    \includegraphics[scale = 0.43]{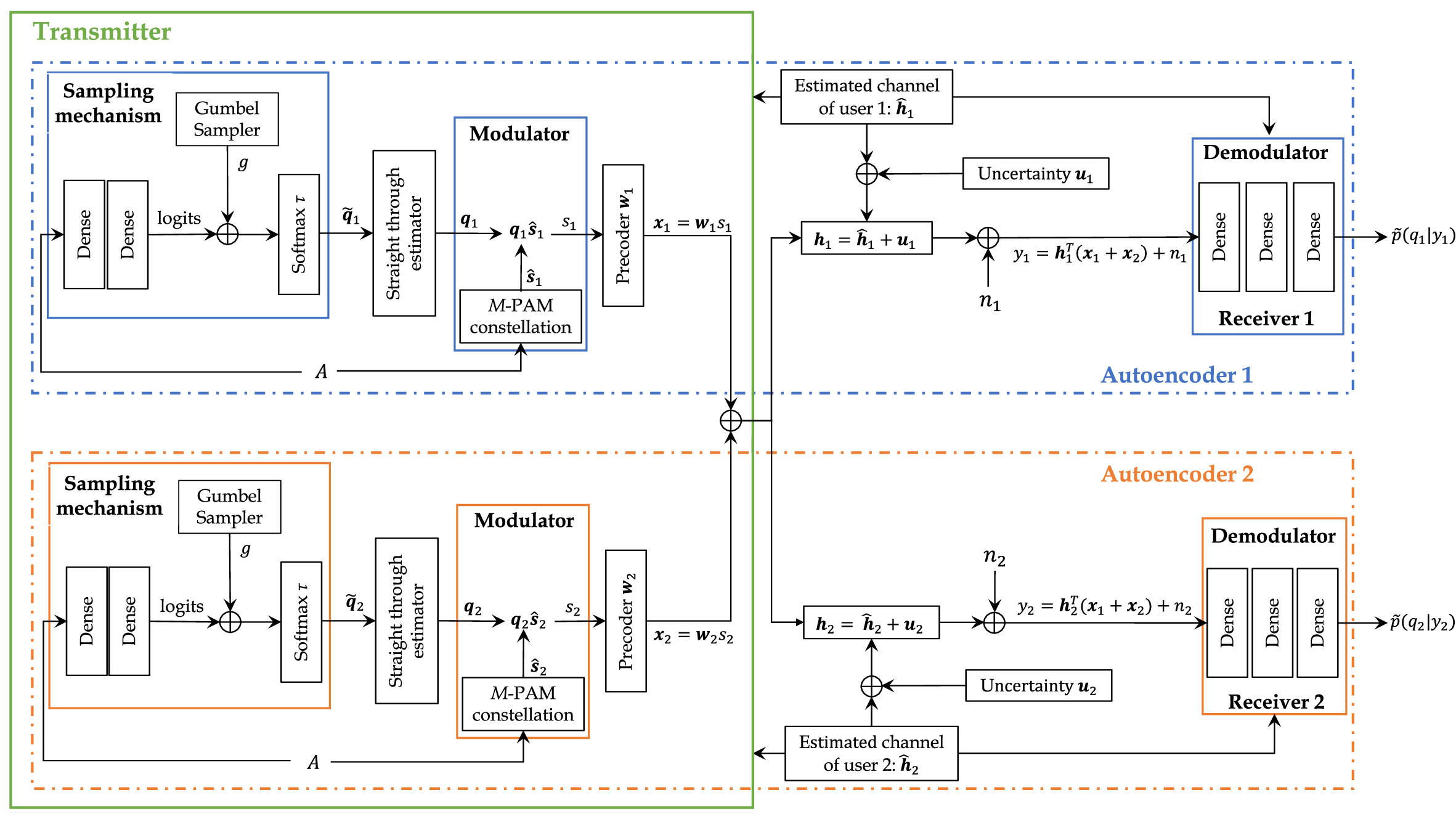}
    \caption{End-to-end system architecture.}
    \label{fig:AE_model} 
    \vspace{-0.5 cm}
\end{figure*}

\subsection{Robust Autoencoder Design}
\label{sec:model_AE}
The idea of using end-to-end learning for physical layer design has been introduced in \cite{intro_DL_com_2017}, where the end-to-end performance of communication systems can be optimized by utilizing a deep neural network (NN) known as AE. By considering the transmitter, channel, and receiver as a single NN, an AE-based communication system allows the transmitter and receiver to be jointly optimized. 

As an example, Fig. \ref{fig:AE_model} depicts the end-to-end communication system architecture for the two-user VLC system, which has one transmitter containing $N_T$ LED luminaires simultaneously serving two users with PCS and precoding. By performing PCS,  the transmit $M$-PAM constellation symbol $s_k \ \in \ \mathbb{S}_k = \{ s_{k,1}, \ s_{k,2}, \ \cdots, \ s_{k,M}\}$ to the $k$-th user is drawn from probabilistically shaped data symbol $q_k \ \in \ \mathbb{Q}_k = \{1, \ \cdots, \ M\}$  corresponding to a parametric probability distribution $p(q_k)$. The data symbol $q_k$ is directly sampled from the trainable sampling mechanism according to $p(q_k)$ and represented as an \textit{one-hot} vector $\mathbf{q}_k$. The transmitter then encodes the constellation symbol $s_k$ to the information-bearing signal $\mathbf{x}_k$ by a precoder $\mathbf{w}_k$, i.e., $\mathbf{x}_k = \mathbf{w}_k s_k$. Afterward, the broadcast signal $\mathbf{x} = \mathbf{x}_1 + \mathbf{x}_2$ is sent to each user. 

At the receiving end, the demodulator of the $k$-th user maps the received signal $y_k$ to a probability vector over the data symbol set to reconstruct the respective transmitted data symbol. For instance, user 1 attempts to reconstruct $q_1$ from observation $y_1 = \mathbf{h}_1^T(\mathbf{x}_1 + \mathbf{x}_2) + n_1$. Because $q_1$ is only contained in the information-bearing signal $\mathbf{x}_1$, we can consider the transceiver block which samples $q_1$, modulates $q_1$ to $s_1$, encodes $s_1$ to $\mathbf{x}_1$, sends $\mathbf{x}_1$ over the channel with channel gain $\mathbf{h}_1$, and reconstructs $q_1$ from $y_1$ as a single AE. Consequently, the design of a general $K$-user VLC broadcast system with PCS can be interpreted as designing $K$ interfering AEs that try to transmit and reconstruct their respective data symbols. Recall from the formulation in  
$\mathbb{P} \mathbf{4}$ that, our objective is finding the optimal precoder $\mathbf{w}_k$'s and optimal distribution of constellation symbol $s_k$ (corresponding to data symbol distribution $p(q_k)$ because of the unique mapping from $q_k$ to $s_k$) to maximize the sum-rate over the channel uncertainty. To achieve this goal, a robust training process is performed to optimize this end-to-end system that contains multiple interfering AEs. As seen in Fig. \ref{fig:AE_model}, each AE contains 3 main components: sampling mechanism, modulator, and demodulator.

\subsubsection{Sampling mechanism}
A major challenge in performing PCS $M$-PAM modulation with machine learning (ML)-based algorithms is training a sampling mechanism for data symbol $q_k \in \mathbb{Q}_k$ drawn from a parametric probability distribution $p_{(k, \theta_{E_k})} (q_k)$ with a trainable parameter $\theta_{E_k}$. Here, $\mathbb{Q}_k = \{1, \ \cdots, \ M\}$ is the event space of the random variable $Q_k$ corresponding to data symbol $q_k$.
As proposed in \cite{Gumbel_max_2012}, the Gumbel-Max trick is a simple and efficient way to sample from $p_{(k, \theta_{E_k})} (q_k)$ as
\begin{align}
    q_k = \argmaxA_{m=1, \ \cdots, \ M} (g_m + \log(p_{(k, \theta_{E_k})}(m))), 
\end{align}
where $g_1, \ \cdots, \ g_M$ are independent and identically distributed (i.i.d) samples drawn from a standard Gumbel (0,1) distribution. However, since the $\argmaxA$ operator is not differentiable, training the optimal symbol distribution $p_{(k, \theta_{E_k})} (q_k)$ by the stochastic gradient descent (SGD) method is not feasible. This issue can be addressed by leveraging the Gumbel-Softmax trick proposed in \cite{Gumbel_softmax_2017}.
The key idea of the Gumbel-Softmax trick is using the softmax function as a continuous, differentiable approximation to $\argmaxA$ function and generating a $M$-dimensional sample vector $\Tilde{\textbf{q}}_k$ with components 
\begin{align}
\label{eqn:temp}
    \Tilde{q}_{k, m} &= \frac{\exp((g_m + \log(p_{(k, \theta_{E_k})} (m))) / \tau)}{\sum_{i = 1}^M \exp((g_{i} + \log(p_{(k, \theta_{E_k})} (i))) / \tau)}, \ m= 1, \ \cdots, \ M, \\
    & \text{and} \ \argmaxA_{m=1, \ \cdots, \ M} \Tilde{q}_{k, m} = q_k,    
\end{align}
where $\tau$ is a positive parameter called the softmax temperature and $\Tilde{\textbf{q}}_k$ is an approximation of the \textit{one-hot} representation of $q_k$. As seen from Fig. \ref{fig:AE_model}, the sampling mechanism is an NN that consists of two dense layers. Since the optimal symbol distribution depends on the ratio of $A/\sigma$ \cite{GM_2016}, $A$  is fed to this NN to generate a continuum of $p_{(k, \theta_{E_k})} (q_k)$ that is dependent on the value of $A$. To do this, the NN firstly generates the logits of $p_{(k, \theta_{E_k})} (q_k)$, which are the unnormalized log probabilities. Then, the Gumbel-Softmax trick is applied to these logits to generate the approximated continuum of $p_{(k, \theta_{E,k})} (q_k)$, which is used for the end-to-end learning. 
Through the learning process, by tuning the trainable parameter $\theta_{E_k}$, the optimal data symbol distribution can be retrieved by applying a softmax activation to the logits. 

\subsubsection{Modulator}
The data symbol is fed into a modulator, which maps each probabilistically shaped data symbol $q_k$ into a $M$-PAM constellation symbol $s_k \in \mathbb{S}_k$ where the constellation symbols are symmetric and equally spaced with maximum amplitude being $A$. Denote the constellation symbol vector is $\mathbf{\widehat{s}}_k = \begin{bmatrix} s_{k,1} \ s_{k,2} \ \cdots \ s_{k,M} \end{bmatrix}$, the constellation symbol is selected by taking the product of the approximated one-hot vector $\Tilde{\textbf{q}}_k$ and $\mathbf{\widehat{s}}_k$. However, because $\Tilde{\textbf{q}}_k$ is only an approximation of the true one-hot vector, the product of $\Tilde{\textbf{q}}_k$ and $\mathbf{\widehat{s}}_k$ is a combination of multiple constellations symbols, causing an infeasibility in choosing the transmitted symbols. Hence, a straight-through estimator is deployed to avoid this problem \cite{straight_estimator_2013}. The key idea of the straight-through estimator is using the true one-hot vector 
for the forward pass and the approximate one-hot vector $\Tilde{\textbf{q}}_k$ for the backward pass. Then, the constellation symbol $s_k$ is precoded by the precoder vector $\mathbf{w}_k$ with trainable parameter $\theta_{w_k}$ to form the information-bearing signal $\mathbf{x}_k = \mathbf{w}_k s_k$ and transmitted through the VLC channel to the users.

\subsubsection{Demodulator}
The demodulator is an NN that consists of three dense layers with the trainable parameter $\theta_{D_k}$, ReLU activations in the first two layers, and softmax activation in the last layer to output the probability vector over the data symbol set $\mathbb{Q}_k$.
\subsubsection{Robust training}
\label{sec: robust_train}
Since the demodulator performs the classification task to reconstruct the respective data symbol $q_k$ from the observation sample $y_k$, the categorical cross-entropy $\mathbb{E}_{q_k, y_k}\{-\log(\Tilde{p}_{\theta_{D_k}}(q_k|y_k))\}$ is commonly used as the loss function \cite{intro_DL_com_2017}. In order to maximize the achievable rate by performing PCS and precoding, the loss function for each AE can be rewritten based on the derivation in \cite{joint_GS_PS_2019} as
\begin{align}
\label{eqn: loss_AE}
    &L_k(\theta_{E_k}, \theta_{w_k}, \theta_{D_k}) =\mathbb{E}_{q_k, y_k}\{-\log(\Tilde{p}_{\theta_{D_k}}(q_k|y_k))\} - h_{\theta_{E_k}}(Q_k) \nonumber \\
    &= \mathbb{E}_y\{\text{D}_\text{KL}(p_{\theta_{E_k}, \theta_{w_k}}(s_k|y_k)||\Tilde{p}_{\theta_{D_k}}(s_k|y_k)\} - \mathbb{I}_{\theta_{E_k}, \theta_{w_k}}(S_k;Y_k), 
\end{align}
where $h_{\theta_{E_k}}(Q_k)$ is the entropy of the random variable $Q_k$ corresponding to the data symbol $q_k$, $\text{D}_\text{KL}(p_{\theta_{E_k}, \theta_{w_k}}(s_k|y_k)||\Tilde{p}_{\theta_{D_k}}(s_k|y_k)\}$ is the Kullback–Leibler (KL) divergence between the true posterior distribution $p_{\theta_{E_k}, \theta_{w_k}}(s_k|y_k)$ and that estimated by the demodulator $\Tilde{p}_{\theta_{D_k}}(s_k|y_k)\}$. And, $\mathbb{I}_{\theta_{E_k}, \theta_{w_k}}(S_k;Y_k)$ is the mutual information of the random variable $S_k$ corresponding to the channel input $s_k$ and the random variable $Y_k$ corresponding to the channel output $y_k$. Therefore, training each AE individually by minimizing $L_k$ corresponds to maximizing $\mathbb{I}(S_k, Y_k)$ and minimizing the KL divergence. Moreover, as described in \cite{joint_GS_PS_2019} and \cite{joint_GS_PS_2020}, the NN implementing the demodulator should be chosen wide and deep enough to learn the posterior distribution with high precision. Assuming a sufficiently wide and deep NN implementing the demodulator, a tight lower bound on the mutual information can be estimated from $L_k$. This avoids learning a constellation where the posterior distribution is well approximated but does not maximize the mutual information.

Recall that the objective of the proposed robust end-to-end learning is to maximize the sum-rate performance through training multiple interfering AEs. Therefore, training each AE independently using its individual loss function defined in \eqref{eqn: loss_AE} may result in degraded performance as only the achievable rate of the corresponding channel is maximized. 
To tackle this issue, for the training process, we define a joint loss function as a weighted sum of all individual losses as 
\begin{align}
\label{eqn:AE_loss}
    L &= \sum_{k=1}^K \rho_k L_k \ \ 
    \text{with} \ \rho_k = \frac{\left\lVert \mathbf{h}_k \right\rVert_2}{\sum_{i=1}^K \left\lVert \mathbf{h}_i \right\rVert_2},
\end{align}
where the weight $\rho_k$ determines the training priority of individual loss  $L_k$ in the joint loss function, i.e., the $k$-th AE with a higher channel gain $\left\lVert \mathbf{h}_k \right\rVert_2$ is trained with a higher priority. It should be noted that aside from \eqref{eqn:AE_loss}, there are several ways of defining a joint loss function, such as giving equal priority to all individual losses (i.e., fixing $\rho_k = \frac{1}{K}$). However, training with equal weight in the joint loss function may not provide a satisfactory performance since it is intuitive that the optimal precoder and constellation distribution to maximize the sum-rate depends on the relativity among users' channel gains. 
This is the main intuition behind the definition in  \eqref{eqn:AE_loss} where the priority of training each AE is proportional to the corresponding channel gain. We note that the proposed joint lost function may not be optimal and that derivation of the optimal one is beyond the scope of the paper. 
The effectiveness of \eqref{eqn:AE_loss} is numerically verified through simulations in the next section. 

Regarding the channel uncertainty, as described in Sec. \ref{sec:4A} and shown in Fig. \ref{fig:AE_model}, based on the estimation from the receivers, the transmitter only knows the estimated channel $\widehat{\mathbf{h}}_k$ and the norm bound of the CSI error $\delta_k$. 
To achieve robustness against the channel uncertainty, the proposed end-to-end learning is trained on varying channel gain vectors, i.e., $\mathbf{h}_k$ by randomly generating the CSI errors $\mathbf{u}_k$ with the norm bound $\delta_k$ in each training batch as
\begin{align}
\label{eqn: uncertainty_channel}
     &\mathbf{h}_k =  \widehat{\mathbf{h}}_k + \mathbf{u}_k, \ \forall{k} = 1, 2, \cdots, K \\\nonumber
     &\text{with } \left\lVert \mathbf{u}_k\right\rVert_2 \ \leq \ \delta_k, \ \delta_k = \alpha \left\lVert \widehat{\mathbf{h}}_k \right\rVert_2.
\end{align}
After training, a joint design of PCS and precoding, which is robust to channel uncertainty, can be learned. 
\begin{figure*}[ht]
    \centering
    \includegraphics[width = 15 cm]{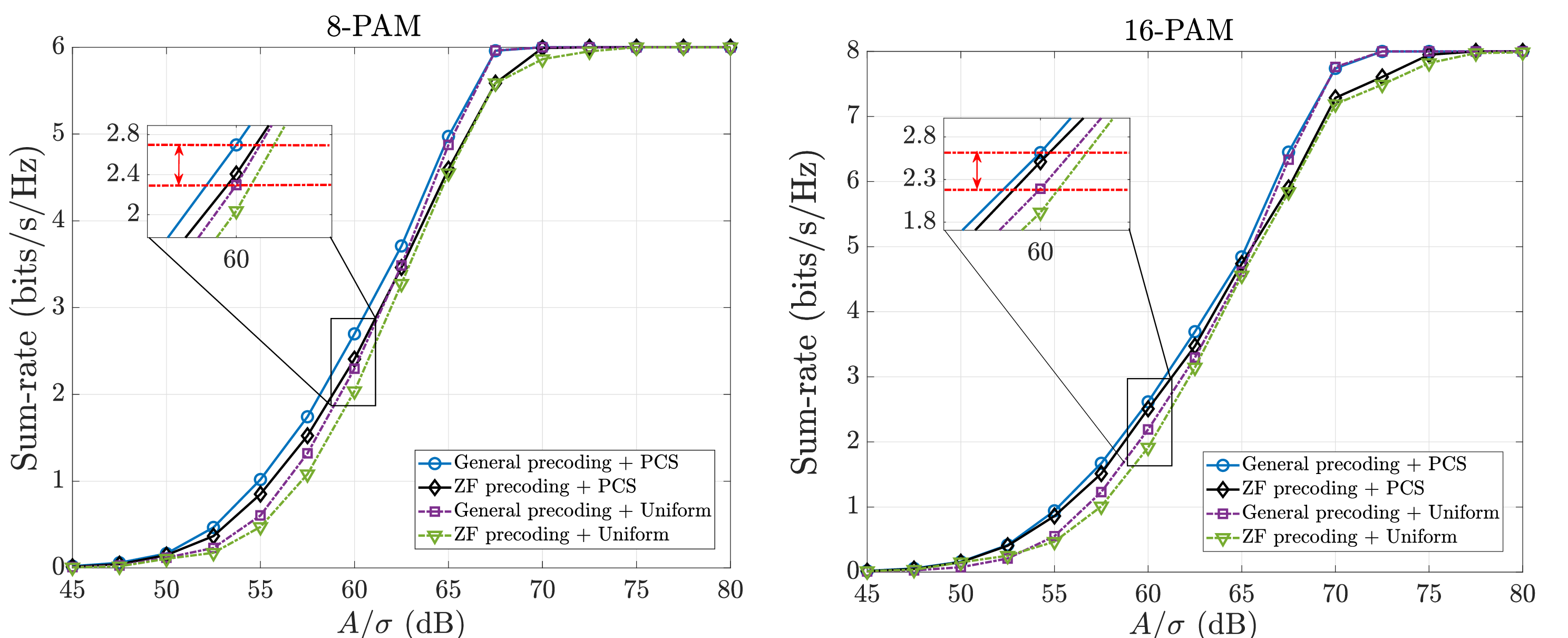}
    \caption{Maximal sum-rate versus $A/\sigma$.}
    \label{fig:SR_A} 
\end{figure*}
\begin{figure*}[!h]
    \begin{subfigure}{1\textwidth}
        \centering
         \includegraphics[width= 15 cm]{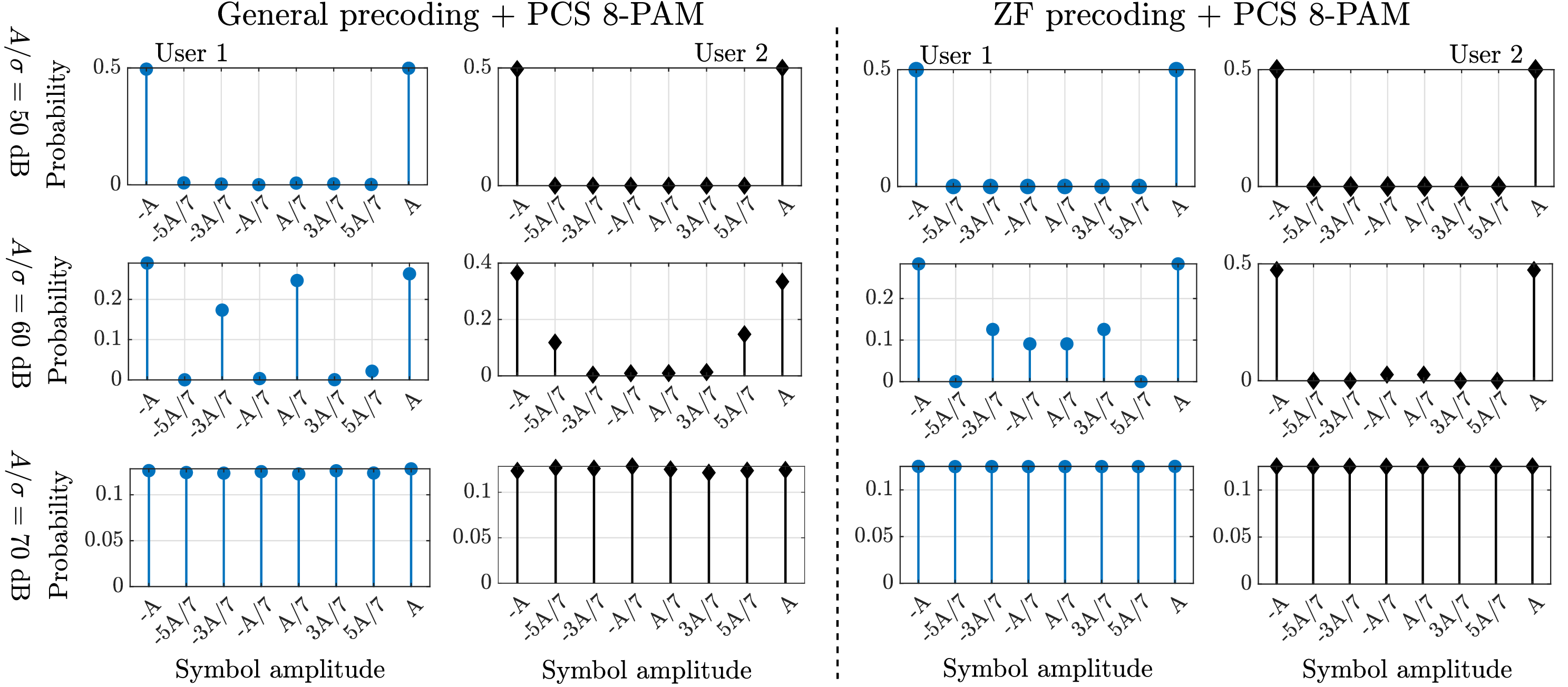}
         \caption{Optimal symbol distributions of PCS 8-PAM with different schemes and $A / \sigma$ values.}
         \label{fig:symbol}
    \end{subfigure}
    \hfill
    \begin{subfigure}{1\textwidth}
        \centering
         \includegraphics[width=15 cm]{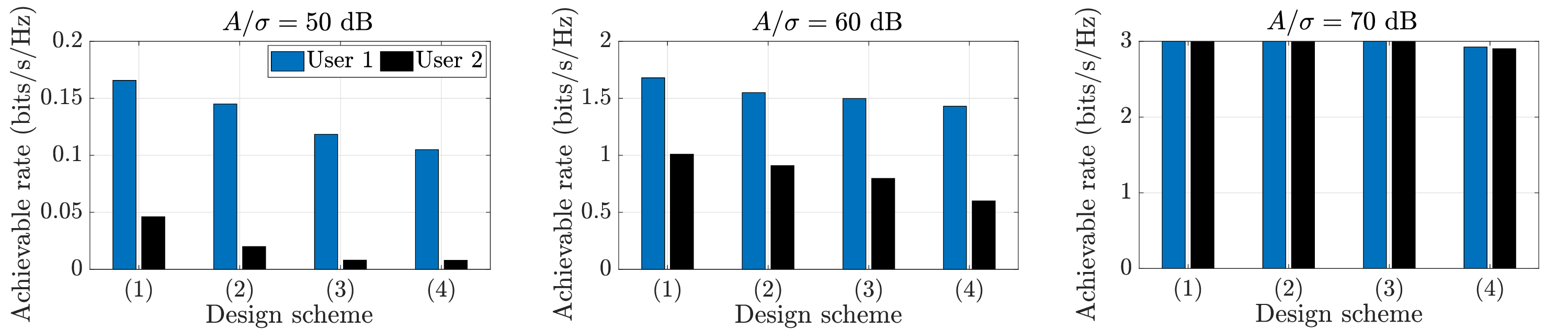}
         \caption{\centering User's achievable rate with different $A / \sigma$ values and schemes:\hspace{\textwidth}
            (1): \centering General precoding + PCS 8-PAM, (2): ZF precoding + PCS 8-PAM\hspace{\textwidth}
            (3): \centering General precoding + Uniform 8-PAM, (4): ZF precoding + Uniform 8-PAM.}
         \label{fig:user_rate}
    \end{subfigure}
    \caption{ Optimal symbol distributions and comparison of user’s achievable rate with different schemes and $A/\sigma$ values.}
\end{figure*}

\section{Simulation Results and Discussions}
\label{sec:results}
This section presents simulation results to demonstrate the effectiveness of the proposed joint design of PCS and precoding for multi-user VLC systems. For simulations, a system consisting of a typical room dimension of $5 \text{m} \times 5 \text{m} \times 3 \text{m}$ in Length $\times$ Width $\times$ Height with four LED luminaries and two users is considered. A 3D Cartesian coordinate system whose origin is the center of the floor is used to specify the positions of the luminaries and users. Then, the four luminaires are positioned at $(\sqrt{2},  \sqrt{2}, 3)$, $(\sqrt{2}, -\sqrt{2}, 3)$, $(-\sqrt{2},  \sqrt{2}, 3)$ and $(-\sqrt{2}, -\sqrt{2}, 3)$ and  positions of user 1 and user 2 are $(0.5, -0.6, 0.5)$ and $(-2.25, 2.2, 0.5)$ respectively. For FA, the variation of attractiveness $\gamma = 1$, attractiveness at zero distance $\beta_0 = 1$, initial random factor $\zeta_0 = 0.9$, population size $N = 120$, number of generations $T = 30$ and penalty constants $\lambda_j = 10^4$ are chosen. Unless otherwise specified, other parameters for the LED configuration and optical receivers are the same as those given in \cite[Table 1]{ZF_2017}.


\subsection{FA- and AO-based design with perfect CSI estimation}
Fig. \ref{fig:SR_A} shows the sum-rate performances of the proposed joint design with PCS and that with uniformly distributed PAM as a function of the peak amplitude-to-noise ratio $A / \sigma$. For both scenarios, the FA-based approach for solving the joint design with general precoding and the AO-based approach for solving that with ZF precoding are compared. It is clearly illustrated that the proposed joint design provides a considerably better sum-rate performance than that with uniform signaling. For example, at $A / \sigma = 60$ dB, the proposed joint design outperforms the uniformly distributed scheme by approximately 0.4 bits/s/Hz ($17.5\%$) for 8-PAM and 0.42 bits/s/Hz ($19.2\%$) for 16-PAM. At the higher $A / \sigma$, however, the rate gap between the proposed design and uniform distributed scheme becomes narrower. This is because, with the optimal precoding, the MUI is significantly eliminated, and as SNR (i.e., $A/\sigma$) increases, it (the MUI) becomes negligible compared to the desired signal. The channel can thus be well approximated as an AWGN channel with an amplitude-constrained signal whose asymptotic (i.e., SNR goes to infinity) optimal input distribution is uniform. 

Fig. \ref{fig:symbol} shows the optimal symbol distributions of PCS 8-PAM for different values of $A / \sigma$. For the case with general precoding, it is evident that as $A / \sigma$ decreases, the number of active symbols (i.e., symbols with non-zero transmission probability) decreases as well. This is because when the channel becomes noisy, transmit symbols should be placed as far as possible so that the impact of noise can be reduced.  
When $A / \sigma$ is sufficiently high (i.e., $A / \sigma = 70$ dB), the number of active symbols increases and the symbol distribution becomes nearly uniform. 
From Fig. \ref{fig:SR_A}, at the high $A / \sigma$ regime, this transition to uniform distribution is evident as the rate gap between the proposed joint design and uniform distributed scheme is diminishing. Also, it is worth noting that, for the design with ZF precoding, the symbol distribution is symmetric around 0. With ZF precoding, the MUI is fully removed, resulting in  AWGN channels with an amplitude-constrained signal. Thus, using $M$-PAM modulation with the symbol interval $[-A, A]$, the achievable rate is maximized by a symmetric symbol distribution with a unique number of active symbols determined by the received SNR that corresponds to $A / \sigma$ value \cite{Capacity_peak_1971, GM_2016}. 
Furthermore, distinct optimal symbol distributions can be observed for different users due to the difference in users' received SNRs. For example, at $A/\sigma = 60$ dB, the symbol distribution of user 1 has more active symbols than user 2 does since the SNR of user 1's channel is higher than that of user's 2 channel, which is confirmed by the higher achievable rate of the former relative to the latter, as shown in Fig. \ref{fig:user_rate}. Comparisons between users' achievable rates obtained from different design schemes in Fig. \ref{fig:user_rate} further emphasize the superiority of joint design over the conventional precoding design with uniform distributed PAM. 
\begin{figure}[t]
    \centering
    \includegraphics[scale = 0.45]{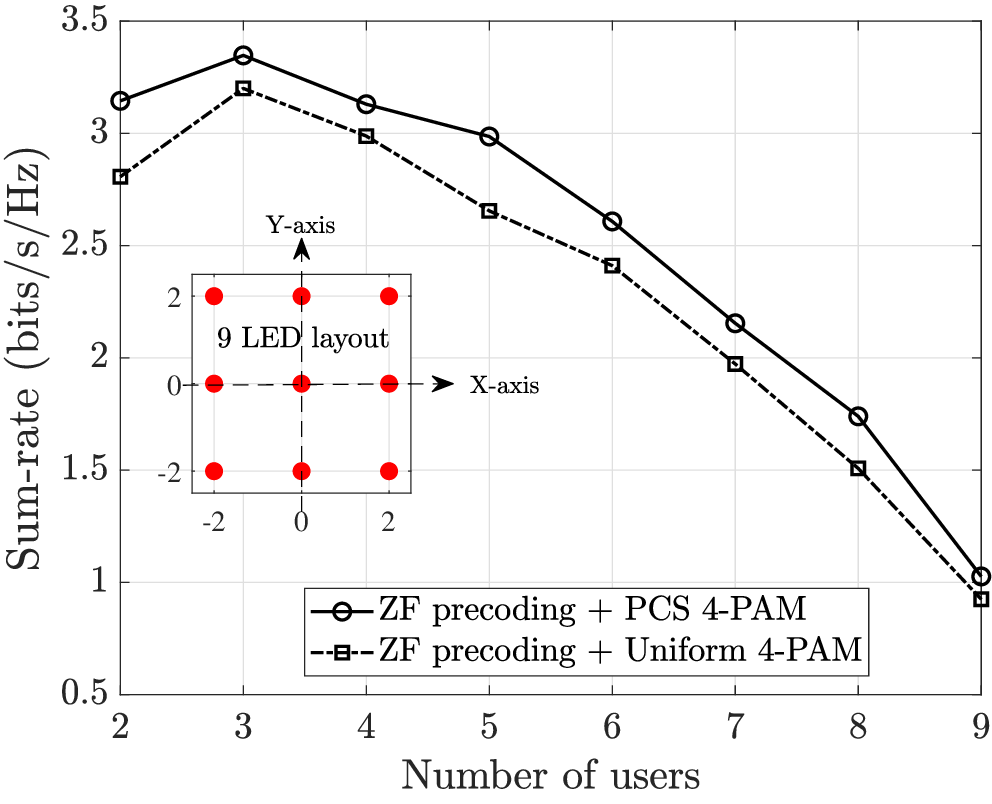}
    \caption{Sum-rate versus the number of users with 9 LED layout, PCS 4-PAM and $A / \sigma = 60$ dB.}
    \label{fig:SR_user}
\end{figure}

In Fig. \ref{fig:SR_user}, we compare the joint design of PCS and ZF precoding with the conventional design of ZF precoding and uniformly distributed PAM for different numbers of user scenarios ranging from 2 to 9. Since the maximum number of users can be supported by ZF precoding equals the number of LED luminaires, 9 luminaries are used for this comparison, and the LED layout is specified in this figure. The results are obtained by averaging from 1000 randomly generated channel realizations. It is observed that the joint design with PCS significantly outperforms the design with uniform signaling. This is because by jointly optimizing the precoding matrix and constellation symbol distributions, the interference between these users is well managed, and the symbol distribution is approximated to capacity-achieving input distribution, leading to better sum-rate performance. Additionally, it is worth noting that when the number of users increases, the sum-rate initially increases and then decreases significantly. This phenomenon is due to the lower degrees of freedom in designing the precoding matrix when the number of users increases (i.e., higher correlation between users' channel vectors). 
\begin{figure}[ht]
    \centering
    \includegraphics[width = 8.9 cm]{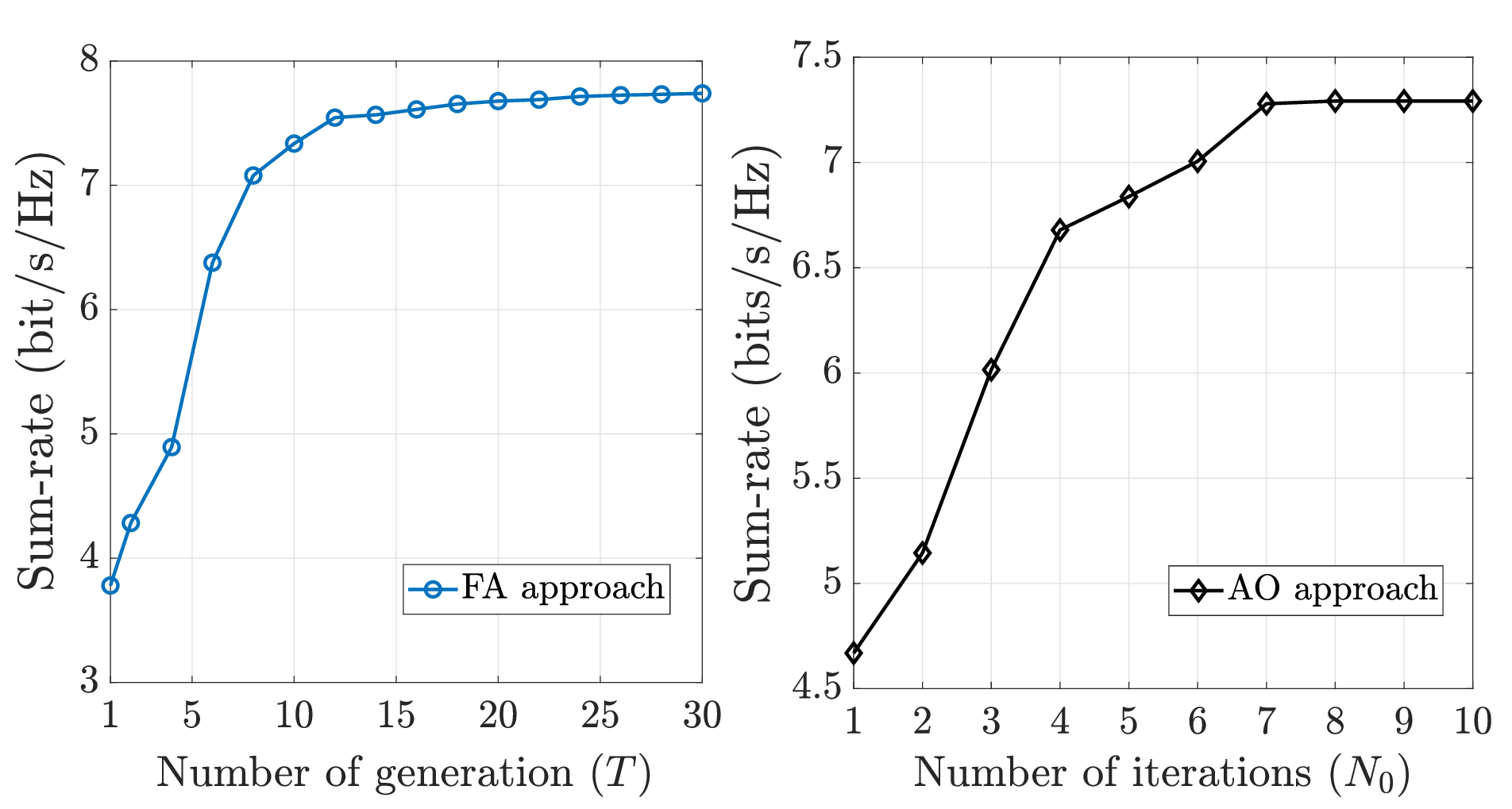}
    \caption{Convergence behaviors of FA and AO approach with PCS 16-PAM and $A / \sigma = 70$ dB.}
    \label{fig:convergence}
\end{figure}

Next, the convergence behaviors of the FA and AO algorithms are presented in Fig. \ref{fig:convergence} with PCS 16-PAM and $A / \sigma = 70$ dB. The proposed FA-based design performs better than the AO-based approach at the expense of increased complexity. In particular, the former attains its optimal solution after 25 to 30 generations, whereas the latter can be solved after only 8 iterations. 
\begin{figure}[t]
    \centering
    \includegraphics[scale = 0.45]{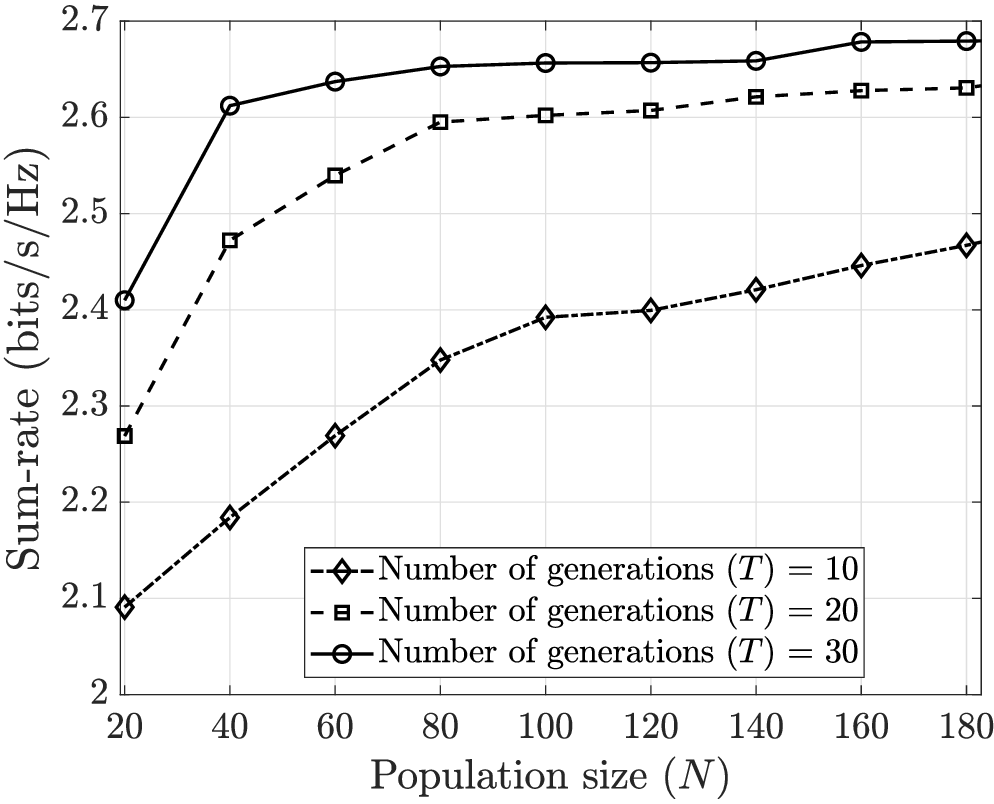}
    \caption{Sum-rate versus the firefly population size with different numbers of generations, PCS 8-PAM and $A / \sigma = 60$ dB.}
    \label{fig:SR_generation}
\end{figure}

Fig. \ref{fig:SR_generation} shows the sum-rate performance obtained from the proposed FA-based design versus the population sizes with different numbers of generations. The observed curves gradually converge when the number of generations increases. This result indicates that increasing the size of the firefly population or extending the search space enables the FA to achieve better solutions.

\subsection{Robust AE-based design with outdated CSI}
This section illustrates the performance of the robust design under channel uncertainty with the proposed end-to-end learning. Training of the end-to-end system presented in Sec. \ref{sec:model_AE} is performed with respect to the loss function $L$ defined in \eqref{eqn:AE_loss} using Tensorflow \cite{tensorflow_2016}. Regarding the NN in each AE, for the sampling mechanism, the first layer is made of 256 units with ReLU activation, and the second layer employs the linear activation with $M$ units ($M$ is the modulation order). For the demodulator, three dense layers are implemented, the first two layers consist of 256 units with ReLU activation, and the last layer is made of $M$ units with softmax activation. Adam SGD optimizer \cite{adam_optimizer_2015} is utilized to train the autoencoders with batches of size 10000, and the learning rate is set to 0.0001. The softmax temperature in \eqref{eqn:temp} is set to 1. The end-to-end system is trained for $A/\sigma$ values ranging from 45 dB to 80 dB. Moreover, regarding the robust training presented in Sec. \ref{sec: robust_train}, giving $\alpha$, the CSI errors $\mathbf{u}_k$ are randomly generated with the norm bound $\delta_k$ to create the varying channel gain vectors, i.e., $\mathbf{h}_k$ in each training batch. 

\begin{figure}[t]
    \centering
    \includegraphics[scale = 0.45]{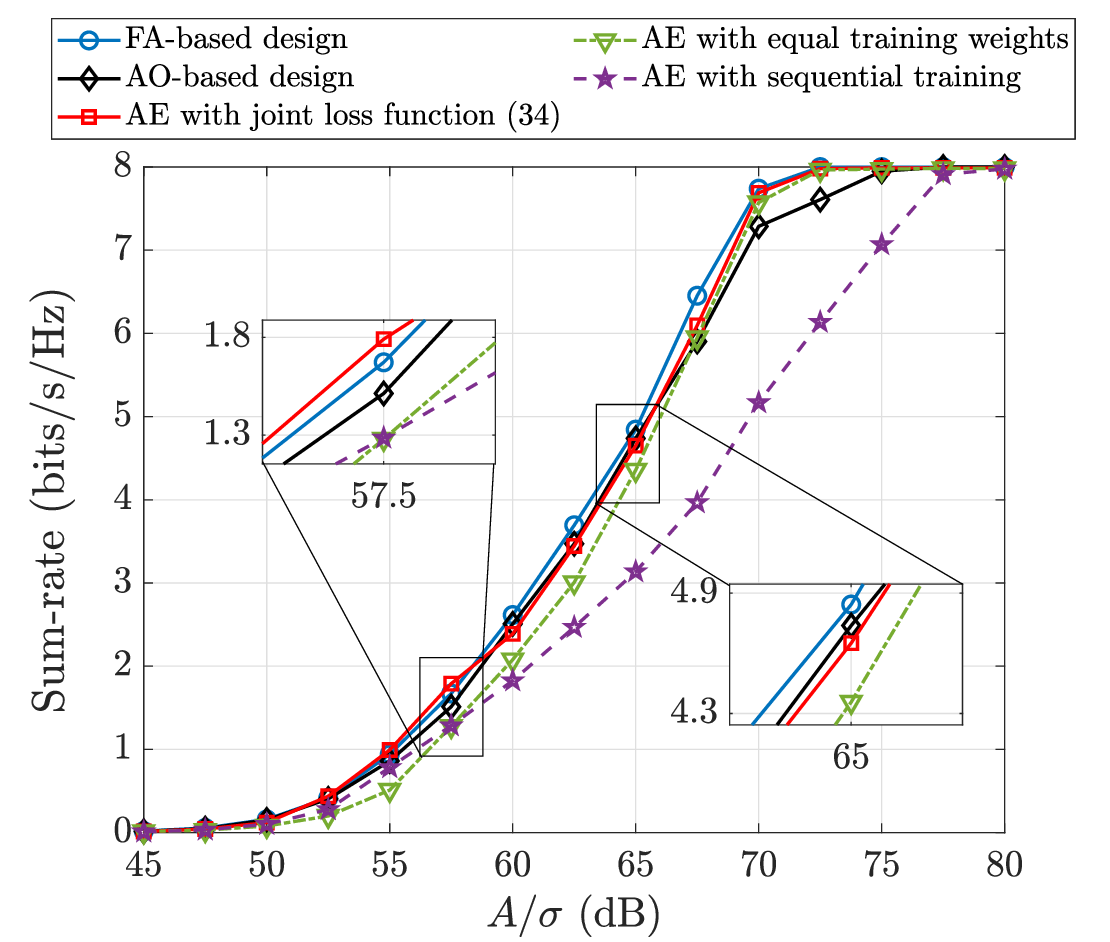}
    \caption{Sum-rate versus $A/\sigma$ for PCS 16-PAM with different designs.}
    \label{fig:SR_AE}
\end{figure}
It is important to note that when the perfect CSI scenario is assumed, the CSI errors $\mathbf{u}_k$ are set 0 in all training batches (i.e., $\mathbf{h}_k =  \widehat{\mathbf{h}}_k$); the learned design is then labeled as ``non-robust". Aside from the FA- and AO-based approaches, we consider this non-robust design as another approach to solving $\mathbb{P} \mathbf{1}$. Therefore, first, the comparison between the performance of the three approaches is depicted in Fig. \ref{fig:SR_AE}.  It is seen that the learned AE-based design can achieve performance very close to that obtained from the FA-based approach, and it is better than the AO-based design in most values of $A/\sigma$. Furthermore, we compare the sum-rate performances obtained by using different loss functions for the proposed robust end-to-end training. Training the end-to-end system, which is a combination of two interfering AE, with the proposed joint loss function in \eqref{eqn:AE_loss} provides a better performance than training with equal priority for each AE, i.e., $\rho_k = 0.5$ in \eqref{eqn:AE_loss}. This improvement comes from the higher priority training for the AE, whose channel gain is better. This result also highlights the ineffectiveness of the training approach where each AE is trained sequentially with its individual loss function \eqref{eqn: loss_AE}. Since the system involves multiple interfering AEs, i.e., $\mathbf{x}_1$ infers the observation $y_2$ at receiver 2 and vice versa, sequentially optimizing each AE may amplify the interference to other AE unintentionally and degrade the sum-rate performance. 
\begin{figure}[t]
    \centering
    \includegraphics[scale = 0.5]{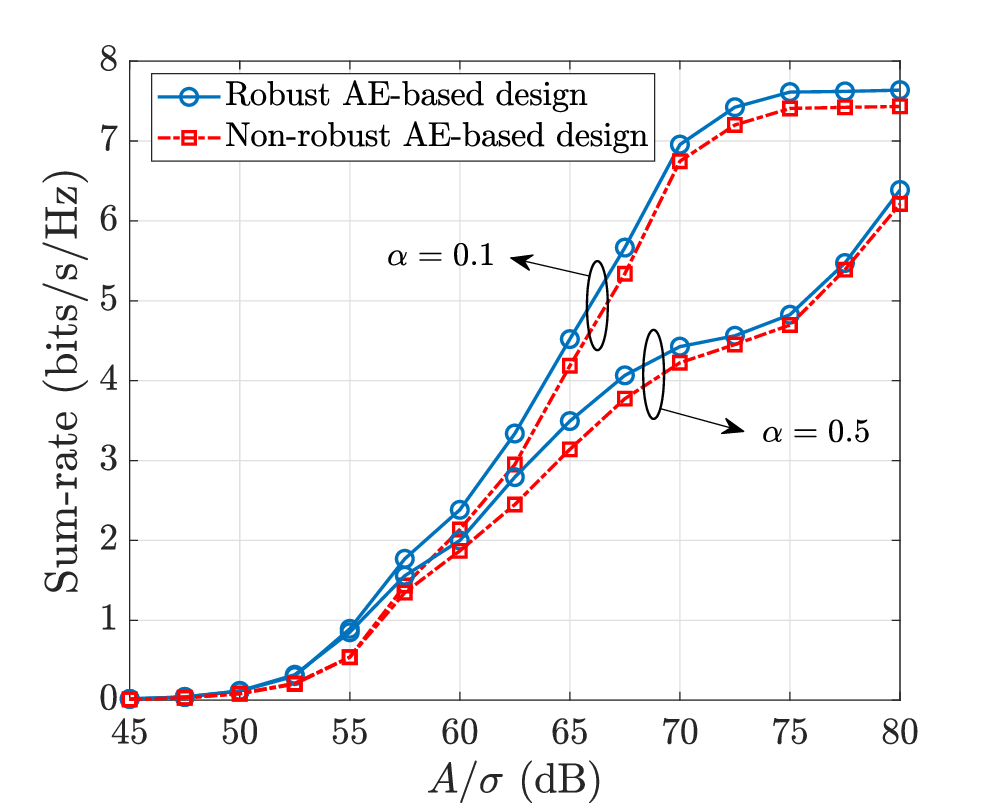}
    \caption{Average sum-rate of the robust and non-robust AE designs versus $A/\sigma$ for PCS 16-PAM with different magnitudes of the CSI error $\alpha$.}
    \label{fig:SR_AE_robust}
\end{figure}

Now, we compare the average sum-rate performance of the learned robust design with the non-robust design. In the robust training process, the magnitude of CSI uncertainty is set to $\alpha=0.5$ in each training batch. For evaluation, with the known channel estimation $\mathbf{\widehat{h}}_k$ and given $\alpha$, the CSI errors $\mathbf{u}_k$ are generated randomly as \eqref{eqn: uncertainty_channel} to form $\mathbf{h}_k$, and the sum-rate performances obtained from robust and non-robust designs are averaged over 2000 trials. As seen in Fig. \ref{fig:SR_AE_robust}, the robust design is superior to its non-robust counterpart. For instance, when $A/\sigma= 65$ dB, the robust design outperforms the non-robust design by approximately $11.2 \%$ when $\alpha = 0.5$ and $8.1 \%$ when $\alpha = 0.1$. By varying the channel gain vectors (i.e., $\mathbf{h}_k = \widehat{\mathbf{h}}_k + \mathbf{u}_k$) during the training process, the robust design can perform better over a range of channel conditions caused by the uncertainty. On the other hand, because the non-robust design is obtained using only the estimated channel vectors, i.e., $\widehat{\mathbf{h}}_k$, the channel uncertainty severely degrades the rate performance. 
\begin{figure}[t]
    \centering
    \includegraphics[scale = 0.5]{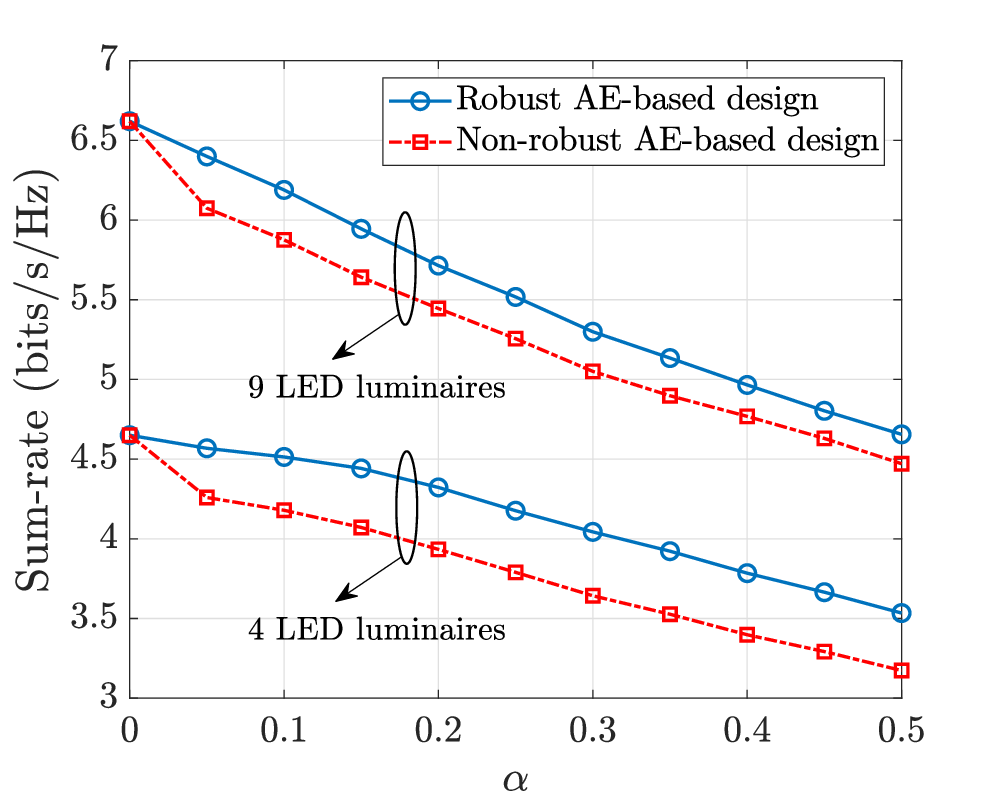}
    \caption{Average sum-rate of the robust and non-robust AE designs versus magnitude of CSI uncertainty $\alpha$ for PCS 16-PAM, $A / \sigma = 65$ dB with different numbers of LED luminaires.}
    \label{fig:SR_AE_robust_noLED}
\end{figure}

Finally, Fig. \ref{fig:SR_AE_robust_noLED} illustrates the average sum-rate performance relative to the magnitudes $\alpha$ of the CSI error for different numbers of LED luminaries (i.e., different sizes of the channel vector). The result again demonstrates that the robust design performs better than the non-robust design when CSI errors occur. For instance, when $\alpha = 0.2$, the improvement obtained from the robust design is approximately $9.1 \%$ when deploying 4 LED luminaires and $5.4 \%$ when deploying 9 LED luminaires. 
Additionally, the result shows that the sum-rate performance increases with the number of LED luminaires. This improvement is attributed to the higher degrees of freedom (DoF) in designing the precoding matrix $\mathbf{W}$ when a larger number of LED transmitters is deployed.

\section{Conclusion}
\label{sec:conclusion}
In this paper, we studied a joint design of PCS and precoding to enhance the sum-rate performance of multi-user VLC broadcast channels subject to peak amplitude constraint. Two different sub-optimal approaches based on FA and AO were presented to solve the design problem. Additionally, considering the channel uncertainty, we also propose a robust design based on the autoencoder concept, an end-to-end learning technique. Numerical simulations verified the superiority of the proposed joint design relative to the conventional design with uniform signaling. The optimal symbol distribution in terms of optical amplitude is also discussed. Furthermore, the advantage of the proposed robust design over the standard ones under uncertain channel conditions is demonstrated by simulation results. For future work, to further enhance the sum-rate performance of multi-user VLC broadcast channels, one may consider a precoding design with joint PCS and GCS. Furthermore, the integration of different multiple access (MA) schemes for multi-user VLC systems, such as non-orthogonal multiple access (NOMA) and rate-splitting multiple access (RSMA) with constellation shaping, could be an interesting research topic. 
\bibliographystyle{IEEEtran}
\bibliography{references}
\end{document}